\documentclass[pdflatex,sn-nature]{sn-jnl}

\usepackage{amsmath}
\usepackage{mathtools}
\usepackage{amssymb}
\usepackage{array}
\usepackage{graphicx}
\usepackage{xr}
\usepackage{setspace}
\usepackage{csquotes}
\usepackage{bibunits}
\usepackage{multirow}
\usepackage{xcolor}
\usepackage{pdflscape}
\usepackage{longtable}
\usepackage{ragged2e}
\usepackage[none]{hyphenat}
\usepackage{booktabs}
\usepackage{rotating}

\usepackage{glossaries}
\newacronym{nsites}{$N_{sites}$}{number of active sites}
\newacronym{m}{$m$}{total mass of electrode}
\newacronym{j}{$j$}{current density}
\newacronym{h}{$h$}{electrode thickness}
\newacronym{phi}{$\phi$}{support:catalyst ratio}
\newacronym{Sm}{$S^{\sigma}_{m}$}{normalized mass sensitivity}
\newacronym{Sphi}{$S^{\sigma}_{\phi}$}{normalized support:catalyst ratio sensitivity}

\defaultbibliographystyle{sn-nature}
\defaultbibliography{cleaned_references}

\begin{document}


\title[Article Title]{Bridging electrode preparation and electrocatalyst performance with physics-based causal AI using Bayesian Structural Causal Models}

\author[1]{\fnm{Evelyna} \sur{Wang}}\email{evelyna.wang@tri.global}
\author[1]{\fnm{Linda} \sur{Hung}}\email{linda.hung@tri.global}
\author*[1]{\fnm{Sam} \sur{Witty}}\email{sam.witty@tri.global}
\author*[1]{\fnm{Michaela} \sur{Burke Stevens}}\email{michaela.stevens@tri.global}
\author*[1]{\fnm{Kevin} \sur{Tran}}\email{kevin.tran@tri.global}

\affil[1]{\orgdiv{Energy \& Materials}, \orgname{Toyota Research Institute}, \orgaddress{\street{4440 El Camino Real}, \city{Los Altos}, \postcode{94022}, \state{CA}, \country{United States of America}}}

\abstract{
State-of-the-art artificial intelligence (AI) and Machine-Learning (ML) tools have not yet enabled rapid design of next-generation materials.  Detailed physical understanding of how material properties affect device performance is required to advance materials development. For example, optimization of ink parameters for electrocatalysts has no known physical mathematical model and thus insights are difficult to translate from material studies to device studies. Herein, we demonstrate how to use the emerging AI tool, \textbf{physics-based structural causal models (SCMs)}, to extract quantitative causative insights from complex heterogeneous electrochemical systems with small (n $<$ 10), but multi-modal datasets. Our SCM quantitatively separates the role that varying the support-to-catalyst ratios and total material loadings plays on catalytic performance. The proof of concept model developed in this work enables root-cause-analysis on the cyclic voltammograms of manganese-antimony oxide oxygen reduction electrocatalysts on Vulcan carbon supports tested in alkaline media using a rotating disc electrode device configuration. Our computational causal analyses quantitatively disentangle how the catalyst performance is affected by the number of active sites versus the thickness of the electrode. To the best of our knowledge, this is the first demonstration of SCMs with physical equations applied to electrochemical materials and their performance. 
}

\keywords{causal reasoning, multiscale modeling, electrocatalysis, oxygen reduction catalysis}


\maketitle

\begin{bibunit}


\section{Introduction}\label{intro} 
The rapid development of industrially relevant new catalysts is limited by the complexity of materials, e.g. catalysts, supports, membranes etc., and their interactions within devices. Fundamental studies focus on \enquote{well defined} catalyst systems to understand isolated catalyst properties~\cite{shinozaki2015oxygen, nie2015recent, fu2019dft, hua2022pdag}. One common method for quickly evaluating electrocatalysts is to use a rotating disc electrode (RDE), which has well studied mass transport, requires minimal material for testing, and is nominally inexpensive~\cite{yarlagadda2017preparation, kocha2017best, shinozaki2018oxygen}. This method is often employed for hydrogen fuel cell catalysts, which catalyze oxygen reduction and hydrogen oxidation reactions (ORR, HOR)~\cite{chen2016evaluation}. The typical protocol for analyzing RDE experiments and comparing catalytic performance is to fit measured voltammograms with standard physical equations, namely the Levich, the Butler-Volmer (Tafel approximation), and the Koutecký-Levich (K-L) equations~\cite{du2014rotating5, du2014rotating7}. The Levich equation is derived from solving a steady-state 1D convective-diffusion equation and describes diffusion of reactants normal to a planar reactive surface. Parameters extracted from the fits are used to assess and compare electrocatalytic activity, including exchange current densities, selectivities, charge transfer coefficients, and limiting current densities. 

Although practical catalysts in fuel cells are typically nanoparticles rather than planar films, performance evaluation of nanoparticle catalysts still often use RDE and Levich analyses due to the accessibility and screening capability of these experiments~\cite{douglin2023high, stamenkovic2007improved}. The planar assumption in the Levich equation, however, presents a method-bias that may result in over- or underestimation of intrinsic electrocatalytic properties of nanoparticles due to differences in electrochemically active surface area, particle faceting, and dispersion~\cite{masa2014koutecky, martin2009rotating}. Furthermore, electrochemical performance of nanoparticulate catalysts may be complicated by electrode design, whereby formulations of inks containing the catalyst, support, and ionomer all affect electrode-microstructure properties~\cite{kreider2022strategies, shinozaki2015oxygenII, deschamps2023rotating, garsany2014analytical, kokhanov2025improving}. Despite many useful improvements to modeling electrocatalysts in RDEs with modifications to the physical equations~\cite{knoche2016film, xu2017building, bonnecaze2007behavior}, as well as applying interpretable machine-learning (ML)~\cite{deo2024interpretable}, it remains challenging to disentangle the effects of intrinsic catalyst properties, other electrode components, or different microstructure on electrochemical performance. As such, thus far, RDE evaluation of catalysts without separating intrinsic and electrode engineering effects remains difficult to translate to larger scale devices (such as membrane electrode assemblies) and use in hydrogen fuel cells~\cite{lazaridis2022capabilities, yarlagadda2017preparation, dull2022alloyed, chen2020Reconsidering, fan2021bridging, zaman2023bridging}. 

New methods are needed to disentangle effects of individual electrode components on experimental performance. One such method is to use Structural Causal Models (SCMs) to quantify root causes and disentangle complex phenomena. SCMs provide the basis for causal AI that can help resolve limitations in conventional data-driven ML tools such as interpretability and explainability, especially given \textit{correlation does not equal causation}~\cite{pearl2009causality, halpern2016actual}. In brief, SCMs provide a mathematical and computational scaffolding for expressing assumed cause-and-effect relationships, plus methods to identify and aggregate unknown relationships and derive causal conclusions. State-of-the art methods in causal modeling not only focus on quantifying effects of one variable on another, but computational advances have also enabled more efficient tools in validating causal structures given data~\cite{spirtes1991algorithm, spirtes2001causation, kalisch2007estimating, chickering2002optimal, zheng2018dags, nazaret2025extremely}. 

SCMs have been employed in the social sciences, where both experimental data and detailed mechanistic knowledge are limited~\cite{klebel2026introduction}. In these settings, practitioners often make use of either \textit{nonparametric structural causal models}, which express causal assumptions solely in terms of causal structure ($x \rightarrow y$ indicates that x influences y) rather than the specific causal mechanisms (\textit{how} x influences y) or \textit{linear structural causal model}, which assume that every causal mechanism is a linear function of its direct causes. Recent work using SCMs for materials science follows in this practice, using nonparametric or linear SCMs and applying causal discovery or validation practices to understand complex experimental datasets connecting synthesis parameters and atomic structures to material properties (e.g., graphene oxide nanoflakes, perovskites)~\cite{motevalli2020understanding,ghosh2024mapping, barakati2025mapping, kalinin2022atomically, tian2025leveraging, seike2025causal}. To advance the application of causal modeling in materials science, we propose to apply the vast pool of physical and chemical insight to causal mechanisms.
In this work we use \enquote{Bayesian parametric structural causal models}~\cite{witty2023bayesian}, which express causal assumptions as probability distributions over a system of one-way equations to \textit{embed physics-based domain knowledge} - e.g., to model RDE experiments, the Levich equation becomes $j_{lim} \leftarrow 0.62nFD^{2/3}\omega^{1/2}\nu^{-1/6}C_{bulk}$, where diffusivity, concentration, and rotation rates \textit{cause} a limiting current density. The inclusion of specified causal mechanisms (inductive bias) enables the use of inference on small experimental datasets as opposed to data-driven methods. While, this framework carries the risk of causal misspecification, we rely on experimental workflows and protocols to help guide the implementation of physical equations; in an RDE experiment, the rotation rate is a control variable that determines a limiting current density rather than the reverse. Along the theme of ML accelerated physics-based models~\cite{aykol2021perspective}, we set out to demonstrate the first integration of well-established physics-based equations into an SCM, thereby enabling systematic causal reasoning (decoupled root cause analysis) on complex electrode design parameters. 

As a method proof-of-concept, we applied structural causal modeling to quantitatively probe the relationship between support ratio, total material mass loading, and electrocatalyst performance based on the experimental dataset from Kreider et al. that probes MnSb\textsubscript{2}O\textsubscript{6} catalysts on Vulcan carbon supports for the ORR in alkaline media using an RDE~\cite{kreider2022strategies}. Our SCM quantitatively connects their small (N$\sim$10) but multi-modal dataset including experimental conditions, compositional information, in situ (double layer capacitance) and ex situ (gas adsorption) surface area measurements, atomic density, density functional theory calculations, and cyclic voltammograms (CVs). Specifically, we use our SCMs to answer the questions: \textit{how does changing the catalyst-to-support ratio and the total material loading affect the total current density measured from voltammetry? Can we decouple the change through electrode thickness versus a change in the total number of active sites?} The application of SCMs with physical mechanisms is a novel approach to answering the broader question, \textit{when changing a material or processing parameter, \underline{why} does performance change?} This work therefore explains in detail the model construction, validation, and causal query analyses methods that are applicable to any electrochemical system. Our methodological results explaining the application of SCMs are detailed in the \nameref{methods}. Overall, this work demonstrates physics-based SCMs and new insights for electrochemical devices that bridge material properties to device performance.


\section{Results}\label{results}

\subsection{Building a structural causal model for ORR}\label{building}

\begin{figure}[htbp!]
    \centering
    \includegraphics[width=0.8\textwidth]{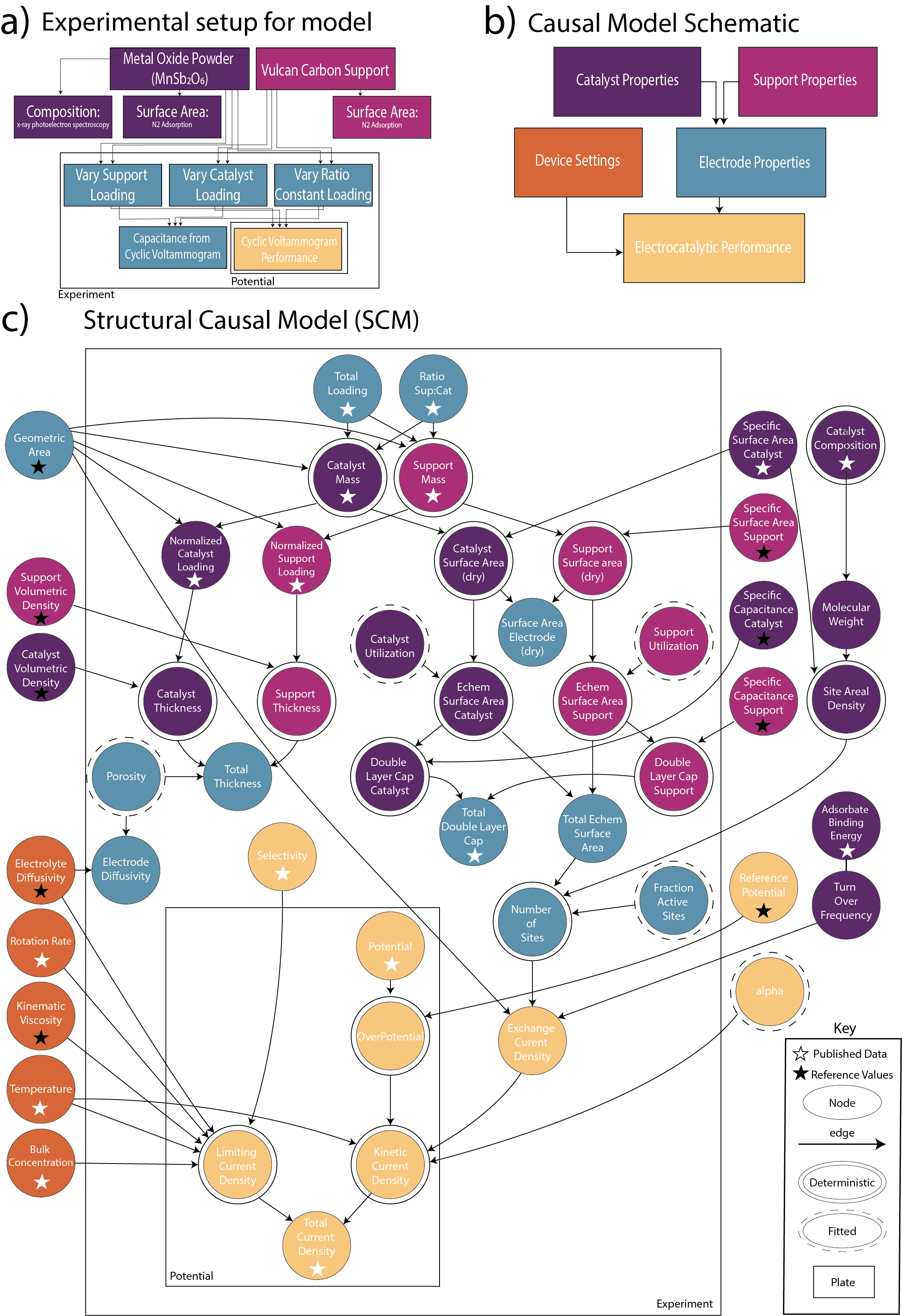}
    \caption{a) A schematic that shows catalyst and support properties \enquote{causing} electrode properties. The electrode and device settings \enquote{cause} the electrocatalytic performance, as modeled by the Levich, Butler-Volmer, and Koutecky-Levich equations, b) a schematic that blocks the experimental design into causal relationships, including multi-modal characterization and experiments varying support and catalyst ratios and cyclic voltammetry measurement, and c) the full classic planar SCM constructed in detail. Nodes labeled with a white star denote experimental data from Kreider et al. and nodes with black star denote literature data or reference values used. Nodes that are deterministically calculated have also been labeled as well as nodes that are \enquote{fitted} during training. There is additional structural information depicted by the nested plating structures, which represent repeated elements in the model, such as the cyclic voltammogram in the potential plate or each individual support ratio/loading in the experimental plate (see \nameref{methods-building SCMs} for further discussion).}
    \label{fig:SCM-schematic}
\end{figure}

We constructed a \textbf{structural causal model (SCM)} of an RDE system to disentangle the effects of composite electrode properties on electrochemical performance. The SCM comprises causal relationships between catalyst properties, electrode properties, device settings, and electrochemical performance and reflects the experimental design from Kreider et al.~\cite{kreider2022strategies}, as shown in Figure~\ref{fig:SCM-schematic}a and ~\ref{fig:SCM-schematic}b. The graphical representation of the full SCM is shown in Figure~\ref{fig:SCM-schematic}c, where all the variables within our model are represented as \textbf{nodes} and directed \textbf{edges} between nodes represent causal relationships. We refer to the model in Figure~\ref{fig:SCM-schematic}c as the classic planar SCM given that the planar assumptions in the Levich equation. The model contains several categories of nodes such as root nodes (nodes without upstream causes) that represent experimental inputs or intrinsic material properties, experimentally observable nodes, and latent variables (parameters that are difficult to measure and therefore fitted during model training. A list of all nodes in the SCM is included in Table~\ref{supp-tab:classic-planar-nodes}.

An example of the causal relationships represented in our SCM is the directed edge connecting the cathodic charge transfer coefficient node to the kinetic current density node ($\alpha \rightarrow$ $j_k$), $\alpha$ is a cause of $j_k$. This cause-and-effect is included in Butler-Volmer mechanism. The kinetic current is then a cause of the measured total current via the Koutecký–Levich equation ($j_k \rightarrow j$). The full SCM as shown in Figure~\ref{fig:SCM-schematic}c assembles the Levich, Butler-Volmer, and the K-L equations together with additional causal relationships between the electrode engineering properties (mass loadings), catalyst properties (adsorbate binding energy, specific surface area), and electrode properties. These include conversion equations (mass and geometric area cause the normalized mass loadings), empirical relationships (Sabatier principle-based volcano relationships - adsorbate binding energy value causes a turnover frequency), as well as estimated relationships (electrochemical surface area catalyst and electrochemical surface area support sum to give Total electrochemical surface area). Table~\ref{supp-tab:classic-planar-nodes} and Figure~\ref{supp-fig:classic-planar-SCM} both show the causal equations plus default values for root nodes. Further discussion on the structure, directionality, and causality when translating physical equations into the causal modeling framework is included in the \nameref{methods}. A \nameref{glossary} is also included. 

\begingroup
\setlength{\tabcolsep}{10pt} 
\renewcommand{\arraystretch}{1.5} 
\begin{table}[!htbp]
  \centering
  \begin{footnotesize} 
    \begin{tabular}{|c|c|c|c|}
      \hline
      \textbf{Experiment} & \textbf{Total material loading}  & \textbf{Support to }  & \textbf{Normalized Catalyst } \\
      \textbf{ID} & (MnSb\textsubscript{2}O\textsubscript{6}  + Vulcan carbon) & \textbf{catalyst ratio}& \textbf{loading} (MnSb\textsubscript{2}O\textsubscript{6})\\
      \hline
      \hline
      A & 0.27 mg$*$cm\textsuperscript{-2} & 0.09 & 0.25 mg$*$cm\textsuperscript{-2} \\
      \hline
      B & 0.31 mg$*$cm\textsuperscript{-2} & 0.2 & 0.25 mg$*$cm\textsuperscript{-2} \\
      \hline
      C & 0.38 mg$*$cm\textsuperscript{-2} & 0.33 & 0.25 mg$*$cm\textsuperscript{-2} \\
      \hline
      D & 0.44 mg$*$cm\textsuperscript{-2} & 0.43 & 0.25 mg$*$cm\textsuperscript{-2} \\
      \hline
      E & 0.125 mg$*$cm\textsuperscript{-2} & 0.2 & 0.1 mg$*$cm\textsuperscript{-2} \\
      \hline
      F & 0.625 mg$*$cm\textsuperscript{-2} & 0.2 & 0.5 mg$*$cm\textsuperscript{-2} \\
      \hline
      G & 1.25 mg$*$cm\textsuperscript{-2} & 0.2 & 1.0 mg$*$cm\textsuperscript{-2} \\
      \hline
      H & 0.625 mg$*$cm\textsuperscript{-2} & 0.09 & 0.57 mg$*$cm\textsuperscript{-2} \\
      \hline
      I & 0.625 mg$*$cm\textsuperscript{-2} & 0.33 & 0.41 mg$*$cm\textsuperscript{-2} \\
      \hline
    \end{tabular}
  \end{footnotesize}
  \caption{Summary of the experiment set we analyzed and used for model training. Each of the experiments will correspond with an experiment plate in the SCM shown in Figure~\ref{fig:SCM-schematic}c; see \nameref{methods-building SCMs} for further discussion.}
  \label{table-experiment}
\end{table}
\endgroup

Once the classic planar SCM was constructed, simulated voltammograms were compared to experimental data to assess preliminary model accuracy, see Figure~\ref{supp-fig:priors-predictive}. We refer to these simulations from the model \textit{before} training with data as \textbf{prior distributions}. The model was then trained on all the experimental data from Kreider et al. (shown in Table~\ref{table-experiment}). During training, the prior distributions of the latent variables in the model are updated in order to better match the distributions of experimentally observable nodes to the measured data values, producing approximate \textbf{posterior distributions}. This fitting process is conceptually similar to what is commonly called \enquote{learning} or \enquote{fitting} parameters in standard ML or physical models, however, in our Bayesian SCMs the result is a set of distributions rather than scalar, vector, or tensor values. Nodes labeled as \enquote{fitted} are attributed with distributions (via kernels; see \nameref{methods}) and updated during training. The list of all applied distributions in the model and rationale are included in Table~\ref{supp-tab:classic-planar-nodes}. To note, the Bayesian SCM framework we implemented incorporates probabilistic methods when modeling uncertainty and training.

\begin{figure}[htbp]
    \centering
    \includegraphics[width=\textwidth]{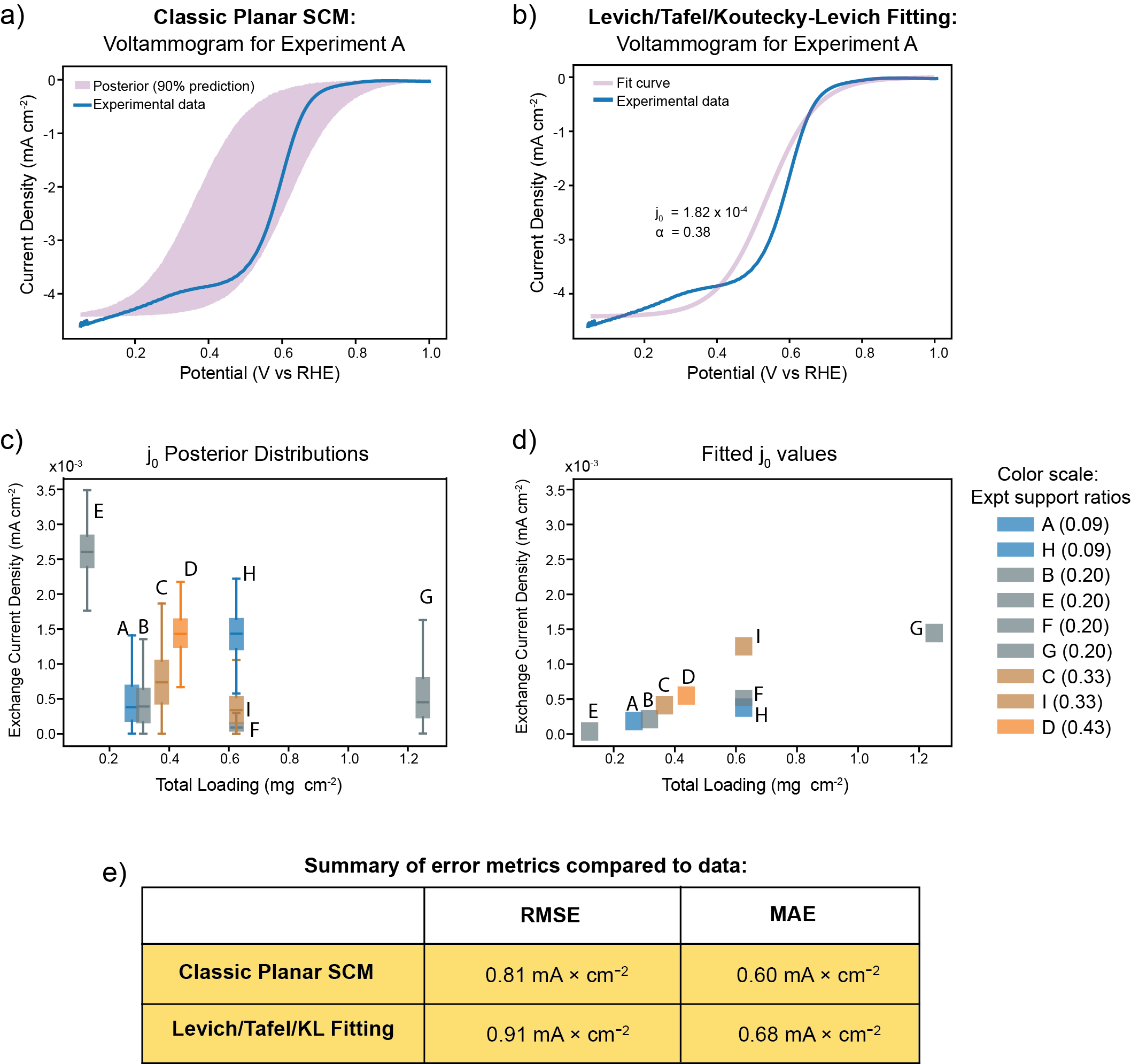}
    \caption{The simulated voltammogram from a) the classic planar model or b) the fit line from Levich/Tafel/KL equation fitting for the reverse sweep of oxygen reduction cyclic voltammogram of MnSb\textsubscript{2}O\textsubscript{6} on Vulcan carbon using an RDE, Experiment A [total loading = 0.27 mg$*$cm\textsuperscript{-2} and support ratio = 0.09]. Resulting posterior distributions for the exchange current density from c) classic planar SCM and resulting fitted exchange current densities from d) conventional experimental data fitting with the Levich/Tafel/KL equation fitting. Both the SCM and conventional fittings were performed with a global charge transfer coefficient across all experiments. Summary of error metrics e) for both the classic planar SCM and the equation fitting for the experimental data indicating poor fits.}
    \label{fig:voltammogram}
\end{figure}

After the training, we compared the data with the simulated voltammograms using the updated model posteriors for latents. The purple band shown in Figure~\ref{fig:voltammogram}a is the posterior distribution for the experiment-specific voltammogram. For comparison, the standard procedure of fitting the Levich, Butler-Volmer, and the K-L equations to experimental data is shown in Figure~\ref{fig:voltammogram}b. 

Figures~\ref{fig:voltammogram}c and~\ref{fig:voltammogram}d show the posterior distributions and extracted fit parameters for the exchange current density ($j_0$) plotted against total mass loading. Although there are some observed trends a (slight downward trend from the SCM and a slight upward trend from the standard equations),the quantitative metrics show large errors (summarized in Figure~\ref{fig:voltammogram}e). The magnitude of these errors undermine the credibility of any observed trends. Additionally, no clear trend exists with support ratio for either fit as indicated by the color legend for the support ratio in Figures~\ref{fig:voltammogram}c and \ref{fig:voltammogram}d. 

The CV posterior predictions and equation fits across the experiments showed poor fitting in the mass transport limiting regime (see Figures~\ref{supp-fig:classic-planar-voltammograms} and \ref{supp-fig:levich-fitting}). This indicates that mass transport effects are not well captured in this set of physical equations (Levich, Butler-Volmer, and Koutecký-Levich) as well as selectivity in the high overpotential regime. Although our classic planar SCM does incorporate electrode parameters along with these physical equations, there is no causal pathway connecting the support ratio or mass loadings to the limiting current, seen graphically in Figure~\ref{fig:SCM-schematic}. As such, we next investigated the validity of an SCM that includes directed edges connecting the electrode thickness to the limiting current density, which required modifications to the Levich equation. 

\subsection{Modeling mass-transport}\label{Mass-transport}

\subsubsection{Modified Levich Equation}\label{ModLevichEqn}

\begin{figure}[htbp]
    \centering
    \includegraphics[width=\textwidth]{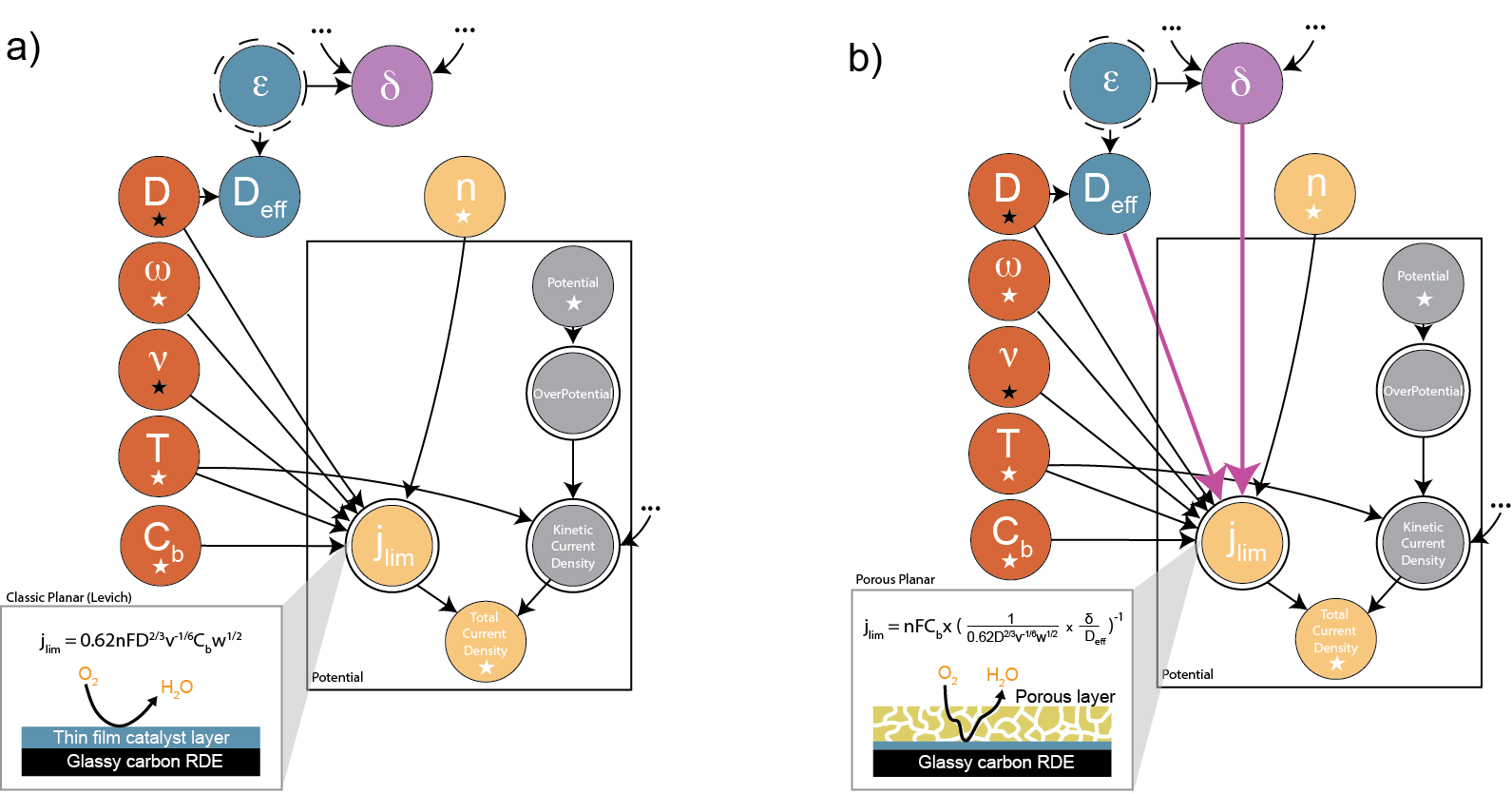}
    \caption{A focused view of the a) classic planar and b) porous planar SCMs highlighting the limiting current density node and its parent nodes. The classic planar uses the Levich equation to model thin film catalyst layers (blue), where the reactants (dissolved O\textsubscript{2}) diffuse across a boundary layer with thickness determined by hydrodynamics. In the porous planar model, the dissolved O\textsubscript{2} hydrodynamic transport is subsequently modified by diffusion through a porous layer (yellow), where the thickness and porosity of this layer dictate the ease of transport (increasing thickness and decreasing porosity both decrease the limiting current density). The equations are shown with the schematics.}
    \label{fig:focused-SCMs}
\end{figure}

To better model the effect of support ratio and total mass loading on mass transport, we constructed a variation of the classic planar SCM that includes both a causal pathway between electrode thickness and the limiting current density as well as a pathway between porosity and the limiting current density. The original Levich equation (Figure~\ref{fig:focused-SCMs}a) is first shown as follows:
\begin{center}
    $j_{lim} \leftarrow 0.62nFD^{2/3}\omega^{1/2}\nu^{-1/6}C_{b}$
\end{center}

Comparatively, the modified Levich equation (Figure~\ref{fig:focused-SCMs}b) is listed below: 
\begin{center}
    $j_{lim} \leftarrow nFC_{b}(\frac{1}{0.62D^{2/3}\omega^{1/2}\nu^{-1/6}}\times \frac{\delta_{tot}}{D_{electrode}})^{-1}$
\end{center}
where the $(\frac{1}{0.62D^{2/3}\omega^{1/2}\nu^{-1/6}})^{-1}$ term represents hydrodynamic transport determined by the RDE rotation rate, identical to the classic Levich equation. The overall transport coefficient includes both the hydrodynamic transport and porous medium diffusive transport, see Figure~\ref{fig:focused-SCMs}b inset. We use $D_{electrode}$ for the diffusivity of O\textsubscript{2} through the porous electrode (Bruggeman relationship: ($D\times \epsilon^{3/2}$), and $\delta_{tot}$ for the electrode thickness. This physical mechanism of hydrodynamics plus diffusion through a porous layer the same as work by Knoche et al with the addition of a causal edge and uncertainty.~\cite{knoche2016film}. We termed this model the porous planar SCM; the full page view is shown in Figure~\ref{supp-fig:scm-porous}. The Butler-Volmer kinetic current equation and the K-L equation modeling the transition from kinetic to mass transport limited regimes remained the same since the kinetics are intrinsic to the material property. Figure~\ref{supp-fig:porous-priors} shows the simulated prior voltammograms from the porous planar SCM. After construction and prior assessment, the prediction performance of the porous planar SCM was quantitatively assessed.

\subsection{Model validation assessment: Prediction performance}\label{validation-prediction}

\begin{figure}[htbp]
    \centering
    \includegraphics[width=\textwidth]{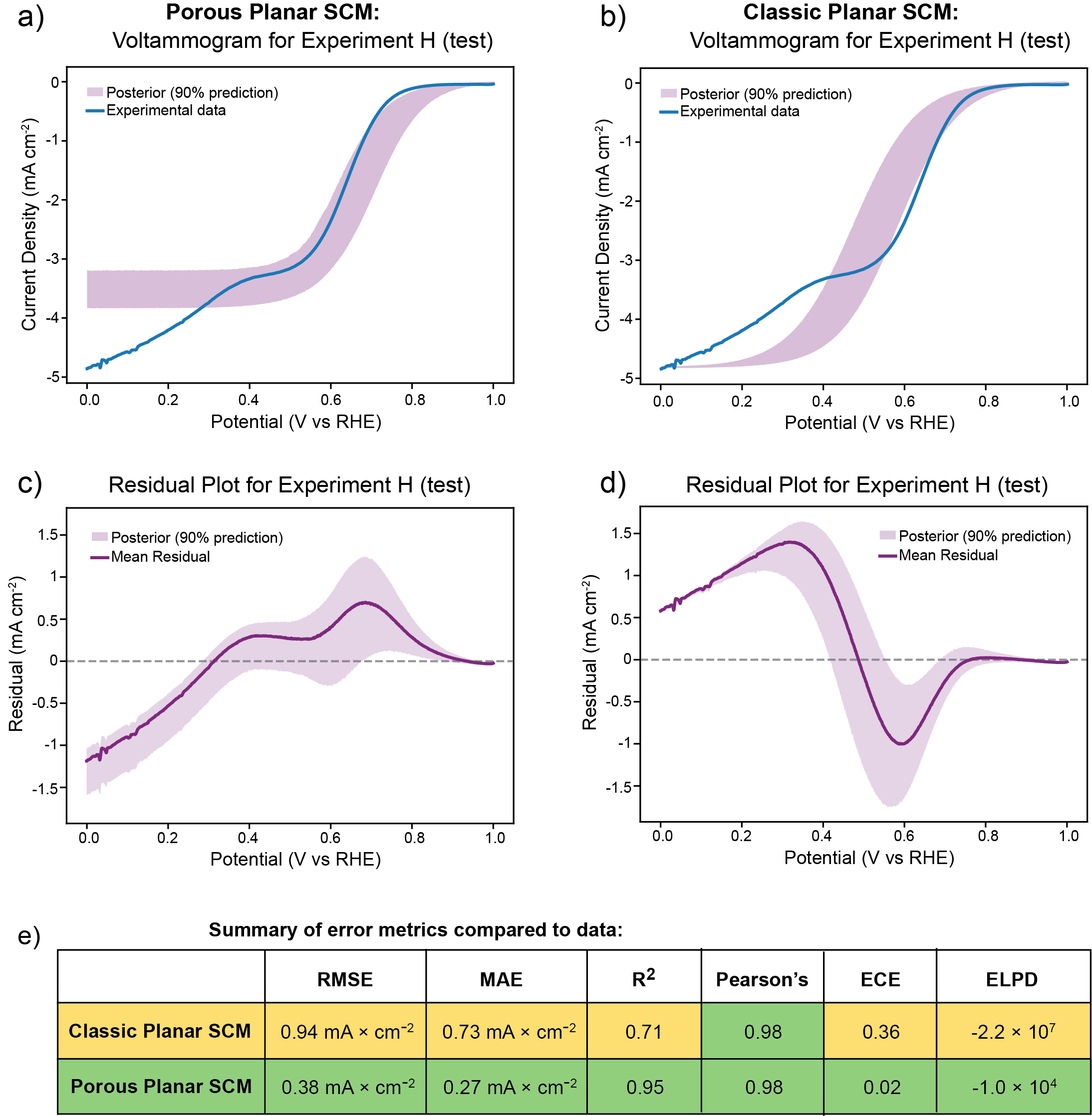}
    \caption{Predictive performance of reverse sweep of oxygen reduction cyclic voltammogram of MnSb\textsubscript{2}O\textsubscript{6} on Vulcan carbon using an RDE for (a) the porous planar SCM, and (b) the classic planar SCM, as evaluated on experiment H from the test set, and (c, d) shows residual plots for experiment H for the porous planar, and classic planar models respectively. The SCMs were trained on the combined training plus validation datasets. (e) Quantitative metrics comparing the predictive performance of the porous planar model and the classic planar models on the test set. The green highlight emphasizes the improvement in performance for the porous planar SCM compared to the classic planar SCM: smaller residual errors (RMSE and MAE), improved goodness of fit (R\textsuperscript{2}), and better calibration (Expected Calibration Error (ECE)) were observed. The Expected Log Predictive Density (ELPD) metric simultaneously evaluates the model’s predictive accuracy and calibration on unseen data~\cite{vehtari2024pareto}: the less negative value for ELPD for porous planar indicates that unseen data is much more likely under the model.
    A 2-sided paired t-test was also performed to assess whether there was a statistically significant difference between the squared residuals of the classic planar and porous planar models.
    Using an $\alpha$ of 0.05 and p-value of $7\cdot10^{-122}$, we rejected the null hypothesis and found that the porous planar model's squared residuals were significantly lower than the classic planar model's.
    This test was performed on the test set with a sample size of 1,000.
    }
    \label{fig:predictions-test}
\end{figure}

Prediction and root-cause analysis using SCMs require different validation procedures. To assess predictive performance, clear separation between test sets and training is necessary, i.e., standard train-validation-test data split as used in conventional ML methods. (For root cause analysis discussed later, we no longer have ground truth and instead train on all available data. See \nameref{workflow} and \nameref{methods-holdouts}.) Figure~\ref{fig:predictions-test}a shows the prior and posterior predictions for the porous planar SCM on the test dataset, experiment H. Compared to the classic planar model, the porous planar SCM showed quantitative improvement across all evaluation metrics as summarized in Figure~\ref{fig:predictions-test}e. Metric details are described in the \nameref{methods}~\cite{tran2020methods, kutner1984applied, vehtari2024pareto}

There remains mismatch in the current density posterior prediction and the experimental data at the mass transport limited regime (Figure~\ref{fig:predictions-test}a,c), indicating further improvements to the porous planar SCM can be made. Figure~\ref{supp-fig:validation-heatmap} shows a summary \enquote{heatmap} pointing to specific latent variable nodes in the model with the most uncertainty. These include fraction active sites and utilization factor nodes. In essence, this highlights that the model relies on the uncertainty in these nodes to better fit the data during training. These nodes and their mechanisms are areas for model improvement. Furthermore, Table~\ref{table:sensitivity-analysis} also shows the prediction performance when sweeping the values of nodes within reasonable physical bounds (e.g., porosity was tested between 0.3 and 0.8); the best prediction performance occurs for most of the default values we originally selected. The ability to both update uncertainty distributions with respect to data and reason on the uncertainty in prediction performance is a major advantage enabled by the Bayesian probabilistic approach we used and part of the quantitative validation of the model before causal inference and interpretation of results.

\subsection{Manual Reasoning on the Porous Planar SCM Posteriors}\label{PorousPlanarPosteiors}

\begin{figure}[htbp]
    \centering
    \includegraphics[width=\textwidth]{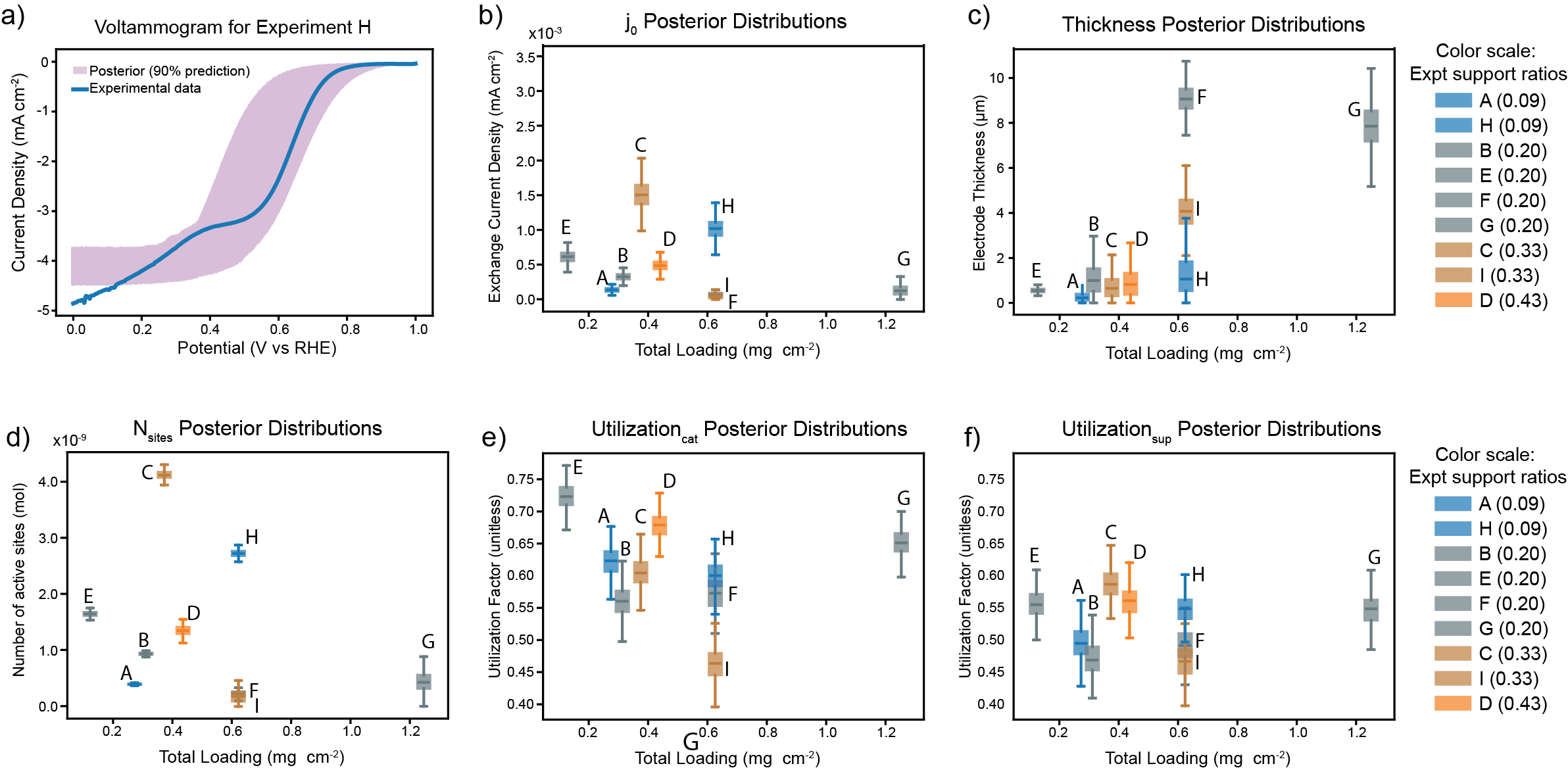}
    \caption{Analyses of the insights from posterior distributions on the reverse sweep of oxygen reduction cyclic voltammogram of MnSb\textsubscript{2}O\textsubscript{6} on Vulcan carbon using an RDE for the porous planar SCM trained on the entire experimental dataset; (a) The posterior prediction voltammogram for experiment H and (b-f) posterior distributions for the various experiments for each of the nodes exchange current density, electrode thickness, number of active sites, catalyst utilization factor, and support utilization factor. The per-node posterior distributions are plotted as a function of increasing total mass loading and colored according to the support ratio in each experiment.}
    \label{fig:posterior}
\end{figure}

While the model was critiqued on predictive performance (with regions identified for improvement), we also analyzed trends in the posterior distributions for the model latents after training on the entire dataset as in a conventional root cause analysis comparing a model's fit parameters. Figure~\ref{fig:posterior}a shows the posterior prediction for experiment H when it was included in the model training. The porous planar SCM posteriors showed good fits to the data even in the mass transport limited regime (Figure\ref{supp-fig:porous-planar-voltammograms} shows the rest of the plots). Figure~\ref{fig:posterior}b shows the posterior distributions for exchange current density node, $j_0$, for each experiment as a function of total loading and colored by support ratio. Rather than a single clear trend, the distributions suggest that even for the same total material loadings (e.g., experiments F, H, I), changing support ratio affects the $j_0$. Figure~\ref{fig:posterior}c shows a reasonable trend in fitted thickness distributions where thickness increases with total loading; although experiment I with a larger support ratio was thinner than experiment F despite the same loading. A direct fitting of the experimental data to the modified Levich/Tafel/K-L set of equations was also performed (Figure~\ref{supp-fig:modified-levich-fitting}), however, the fit gave nonphysical thicknesses and unrealistic trends where the lowest loading experiment had the thickest electrode.

To investigate the root cause of support ratio and loadings on exchange current density, we plotted the posterior distributions of nodes \textit{causally upstream} of $j_0$, including $N_{sites}$ and utilization factor nodes. From the posterior distributions in Figures~\ref{fig:posterior}d-f, no clear trends could be observed with respect to loading or support ratio. Therefore, while we began to investigate the effect of support ratio and material loading on catalyst performance by applying critical thinking on the trained model outputs, trends from this dataset remain challenging to disentangle when using conventional model analysis techniques. As such, we next focused on applying systematic and algorithmic causal reasoning on the SCM and quantified the effects of support ratio and total loading through different causal pathways. 

\subsection{Decoupling support ratio and material loading using mediation analysis}

Our findings from this section are two-fold:  (1) the development of systematic and algorithmic causal reasoning methods for physics-based SCMs, and (2) the quantified answer to the question, \textit{how does changing the catalyst-to-support ratio and the total material loading affect the total current density measured from voltammetry through a change in electrode thickness versus a change in the total number of active sites?} To ask the causal question and perform quantitative reasoning, we developed and applied dedicated software for querying SCMs (see \nameref{methods-causalprobprog}) to the porous planar SCM. Answering additional questions is out of scope for this work as they may require the addition of new nodes or causal pathways in the models; the scientific software tools we developed in this manuscript however, can still be used for these future model updates and questions(\nameref{code}).

\begin{figure}[htbp]
    \centering
    \includegraphics[width=\textwidth]{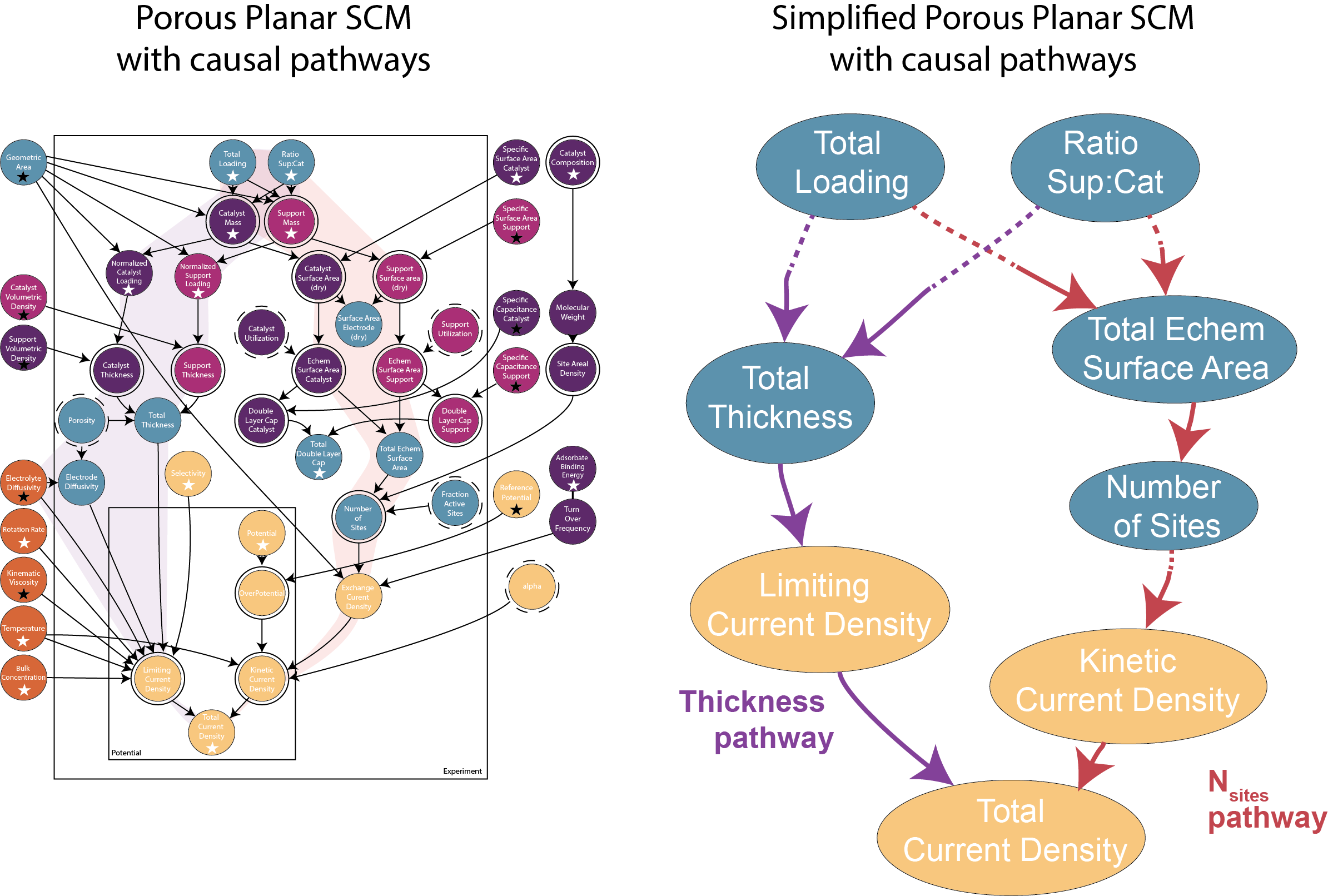}
    \caption{Highlighted visualizations of the causal pathways in the porous planar model; (a) condensed view of the full porous planar SCM highlighted pathways with the purple highlight indicating the mass transport pathway via thickness, and red highlighting the kinetics pathway via number of sites and (b) a simplified view of the porous planar model with the key nodes for each pathway.}
    \label{fig:causal-pathway}
\end{figure}

The porous planar SCM contains two pathways where changing the support ratio or total mass can affect the total current density. The causal pathways are highlighted in Figure~\ref{fig:causal-pathway}; the purple pathway indicates a mass transport cause-effect from changing electrode thickness, and the red pathway is a kinetic cause-effect from changing the number of active sites. For instance, increasing the total mass loading can increase the total thickness of the electrode and further constrain the limiting current density; or increasing total mass loading can increase the number of active sites available and thus increase the exchange current density. Increasing the support ratio can also increase the thickness of the electrode. 

\begin{figure}[htbp]
    \centering
    \includegraphics[width=0.95\textwidth]{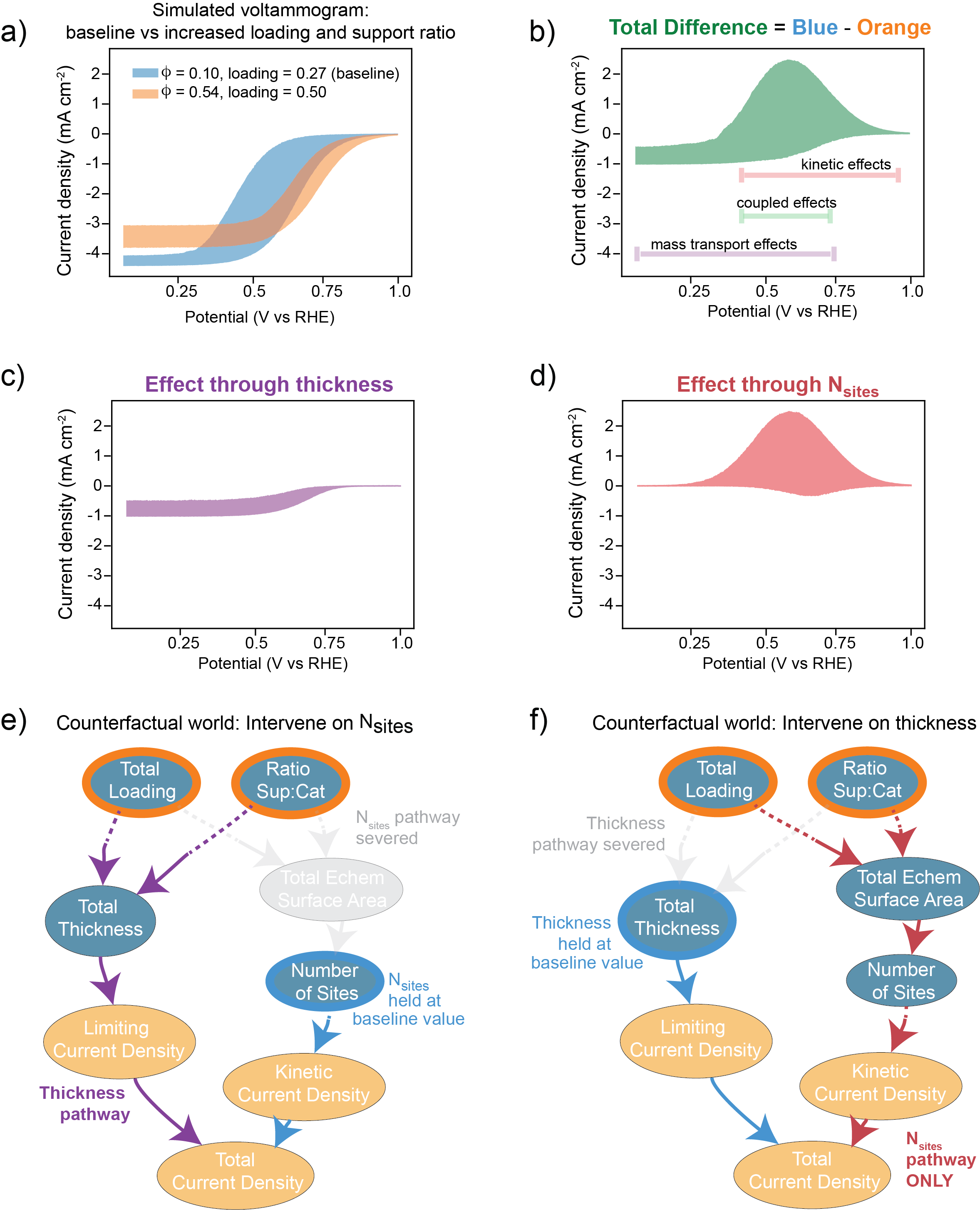}
    \caption{Mediation analysis results of oxygen reduction cyclic voltammogram of MnSb\textsubscript{2}O\textsubscript{6} on Vulcan carbon using an RDE with (a) the simulated voltammograms for the blue baseline experiment (support ratio = 0.10, total loading = 0.27 mg$*$cm\textsuperscript{-2}) and orange new experiment (support ratio = 0.54, total loading = 0.50 mg$*$cm\textsuperscript{-2}), (b) the difference between the blue and orange, (c) the effect of changing the support ratio and total loading through the electrode thickness pathway only showing decrease in performance (negative effect a.k.a. decreased current density magnitude) and (d) the effect of changing the support ratio and total loading through the $N_{sites}$ pathway only showing an increase in performance (positive effect a.k.a. increased current density magnitude). The counterfactual instantiations that enabled the isolated effect through each pathway are shown in (e) and (f), respectively, where the structural function of the thickness or $N_{sites}$ has been replaced. These can be considered simulations of \enquote{impossible control experiments}}.
    \label{fig:mediation}
\end{figure}

To interrogate the two causal pathways, the trained porous planar SCM was used to systematically simulate different experiments and quantitatively compare between them. First, we simulated a baseline experiment with support ratio of 0.09 and total loading of 0.27 mg$*$cm\textsuperscript{-2} (analogous to experiment A from Kreider et al., where experiment A was chosen as the benchmark due to its comparatively small loading and support ratio relative to the other experiments Table~\ref{table-experiment}); the simulated voltammogram distribution is shown in blue in Figure~\ref{fig:mediation}a. We then simulated an experiment with support ratio 0.54 and total loading of 0.5 mg$*$cm\textsuperscript{-2}, shown in orange. The difference between the blue and the orange distributions is shown in Figure~\ref{fig:mediation}b. 

Quantifying the contributions from each of the causal pathways required additional simulations. In a third simulation, we constrained the value for the $N_{sites}$ to its value from the baseline experiment. This severs the causal pathway upstream of the $N_{sites}$ (highlighted blue, Figure~\ref{fig:mediation}e) such that changing the support ratio and the total loading to the values in orange (support ratio 0.54 and total loading of 0.5 mg$*$cm\textsuperscript{-2}) can \textit{only} cause a change in the total current density through the thickness pathway (purple arrows). In causal reasoning terms, we instantiated a \textbf{counterfactual world}, in which we \textbf{intervened} on the total loading and support ratio \textit{as well as the $N_{sites}$}. The counterfactual world of \enquote{effect through thickness only} is the instantiation shown in Figure~\ref{fig:mediation}e, and can be considered an \enquote{impossible control experiment}. The delta between total current density simulated from the counterfactual world in Figure~\ref{fig:mediation}e and the baseline experiment results in Figure~\ref{fig:mediation}c. Similarly, in a fourth simulation, we constrained the value for the thickness to its baseline value, i.e., the counterfactual world shown in Figure~\ref{fig:mediation}f, and the quantified \enquote{effect through $N_{sites}$ only} is shown in Figure~\ref{fig:mediation}d.

The effect through thickness (Figure~\ref{fig:mediation}c) showed that increasing the support ratio to 0.54 and increasing the total loading 0.50 mg$*$cm\textsuperscript{-2} while keeping the $N_{sites}$ at its value in the baseline caused a decrease in total current [magnitude]. Higher support ratios and loadings increased the thickness, causing a decreased limiting current value. In contrast, the effect through $N_{sites}$ showed that increasing the support ratio to 0.54 and increasing the total loading 0.50 mg$*$cm\textsuperscript{-2} while keeping the thickness at its value in the baseline resulted in increased current density [magnitude] (Figure~\ref{fig:mediation}d). This procedure of comparing simulated, interventional control experiments is called \textbf{counterfactual reasoning}, and using counterfactual reasoning to quantify the path-specific effects is called \textbf{mediation analysis}. Further discussion on the mediation analysis methods are in the \nameref{methods-mediation}.

\begin{figure}[htbp]
    \centering
    \includegraphics[width=\textwidth]{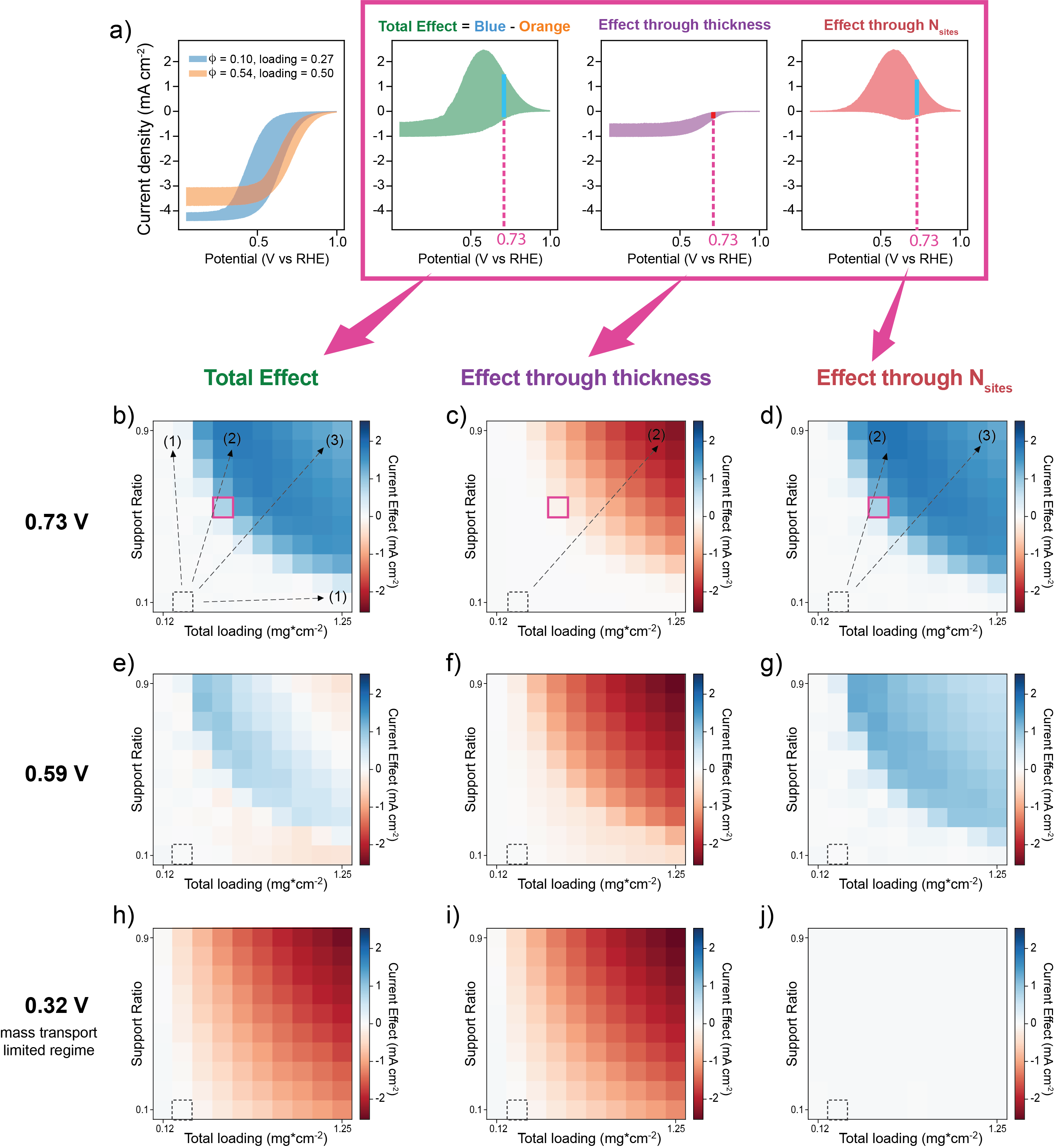}
    \caption{Heatmaps showing the systematic mediation analyses comparing simulated experiments for 100 combinations of support ratio (0.1-0.9) and total loadings (0.12$-$1.25mg$*$cm\textsuperscript{-2}) with reference to the baseline support ratio = 0.10, total loading = 0.27 mg$*$cm\textsuperscript{-2}. Each square in the grid represents the results at a specific potential for a single mediation analysis as explained in Figure~\ref{fig:mediation} and depicted again in (a). A heatmap is generated per-potential and for the (b,e,h) total effect, (c,f,i) effect through $N_{sites}$ via the counterfactual world in Figure~\ref{fig:mediation}e, and (d,g,j) effect through thickness via the counterfactual world in Figure~\ref{fig:mediation}f. The black dotted square highlights the baseline experiment. The color scale indicates a positive effect on current density magnitude (increased performance) with blue shades and a negative effect on current density (decreased performance) with red shades. The pink square highlights the effects at 0.73 V for the support ratio = 0.54 and total loading = 0.5 mg$*$cm\textsuperscript{-2} example, corresponding to the pink dashed lines in (a). The black arrows indicates different trends observed compared to the baseline experiment: 1) no trend, 2) monotonic increase or decrease, and 3) non-monotonic behavior. All potentials referenced to RHE. We also emphasize the method development enabling the application of SCMs and causal reasoning to this electrocatalyst system shown in this figure as part of our results.}
    \label{fig:heatmaps}
\end{figure}

The mediation analysis was systematically performed for 100 combinations of support ratio (0.1-0.9) and total loadings (0.12$-$1.25mg$*$cm\textsuperscript{-2}) with reference to the baseline, as discussed in Figures~\ref{fig:mediation}. The results of the 100 mediation analyses were compared at different potentials, as shown in the per-potential and per-effect heatmaps in Figure~\ref{fig:heatmaps}b-j, where each square in the grid is extracted from a separate set of the simulations discussed in the previous paragraphs. We emphasize here again that the results from the mediation analysis are a demonstration of this new methodology to deconvolute coupled effects such as increased $N_{sites}$ and thickness caused by increasing support ratios and total material loadings. The development of the porous planar SCM and application of probabilistic programming enabled this preliminary effort into root cause analysis in electrochemical devices. 

To discuss the mediation heatmap results in detail, we first look at the 0.73 V heatmap for total effect (Figure~\ref{fig:heatmaps}b). The pink highlighted square indicates the total effect on current density caused by increasing the support ratio to 0.54 and total loading to 0.5 mg$*$cm\textsuperscript{-2} compared to the baseline experiment (dotted black square, support ratio = 0.10 and total loading = 0.27 mg$*$cm\textsuperscript{-2}). A positive current effect at the pink square is observed, as emphasized by the blue color scale, i.e., the current density magnitude increased at 0.73 V, corresponding to increased performance. We note three main trends in Figure~\ref{fig:heatmaps}b when changing support ratio and total loading: starting at the baseline dotted black square and following the black dotted arrows, we observe (1) negligible changes, (2) monotonic increase, or (3) non-monotonic changes in total current. The non-monotonic trend indicates convoluted causes that we begin to disentangle using the mediated effect results from the two causal pathways, $N_{sites}$ and thickness.

The quantified effect through thickness and effect through $N_{sites}$ at 0.73 V heatmaps are plotted in Figures~\ref{fig:heatmaps}c and~\ref{fig:heatmaps}d, respectively. A monotonic increase in performance is observed when increasing to high ratios and loadings via the thickness pathway (Figure~\ref{fig:heatmaps}c, dotted arrow (2)). This trend indicates that at high support ratios and loadings, thick electrodes will decrease catalytic performance when number of sites is held at baseline. Intuitively, this aligns with the non-monotonic change in the total effect, Figure~\ref{fig:heatmaps}b dotted arrow (3), to indicate a trade-off between more material (more active sites) with thicker electrodes (decreased transport). The mediated effect through thickness decouples part of the the non-monotonic trend observed in the total effect. 

In contrast, the effect through $N_{sites}$ at 0.73 V, Figure~\ref{fig:heatmaps}d, shows a monotonic increase in performance is observed when increasing from the baseline to moderately low total loadings and high support ratios via the $N_{sites}$ pathway (dotted arrow (2)) but a non-monotonic increase is observed when increasing to high support ratios and high loadings, (dotted arrow (3)). This means that high loadings and high support ratios beyond $\sim$0.54 do not increase $N_{sites}$ as much as increasing to moderate loadings with moderate support ratios when thickness is held at the baseline value. There may be an optimization of $N_{sites}$ when changing support ratio and total loading that is \textit{independent} of the electrode thickness effect, discussed later.

Next, looking at the total effect heatmap for 0.59 V (Figure~\ref{fig:heatmaps}e), we observed similar convoluted non-monotonic trends. Since the 0.59 V is around the kinetic/mass transport regime transition, the effects of both thickness and $N_{sites}$ are more pronounced than at 0.73 V. For the heatmaps at 0.32 V (Figure~\ref{fig:heatmaps}h-j), which is strongly in the mass transport limited regime, only a decrease in performance due to increased electrode thickness is observed. Together, the heatmaps in Figure~\ref{fig:heatmaps} emphasize the thickness-$N_{sites}$ trade-off on performance and also point to the additional convoluted trends in $N_{sites}$ caused by support ratios/total loadings.

The mediation analysis with the porous planar SCM reveals the observed causes only after training on the experimental data. In other words, these simulations from the trained model reflect what can be learned from the experimental dataset. Notably, the heatmaps from an untrained model (Figure~\ref{supp-fig:mediation-heatmaps-untrained}) only show the trend of poor mass transport at mass transport limited regimes. Figure~\ref{supp-fig:mediation-heatmaps-others} shows the same set of heatmaps but referencing different experiment baselines. Similar trends are observed that also highlight the trade-offs between decreased mass transport though electrode thickness and increased kinetics through $N_{sites}$ when at high loadings and support ratios as well as the possibility of some $N_{sites}$ optimization. Maximizing the insights from the SCMs therefore requires understanding the experimental data and questions, building a representative model, and training the model, which we discuss as a proposed workflow.

\subsection{Structural Causal Modeling Workflow}\label{workflow}

\begin{figure}[htbp]
    \centering
    \includegraphics[width=\textwidth]{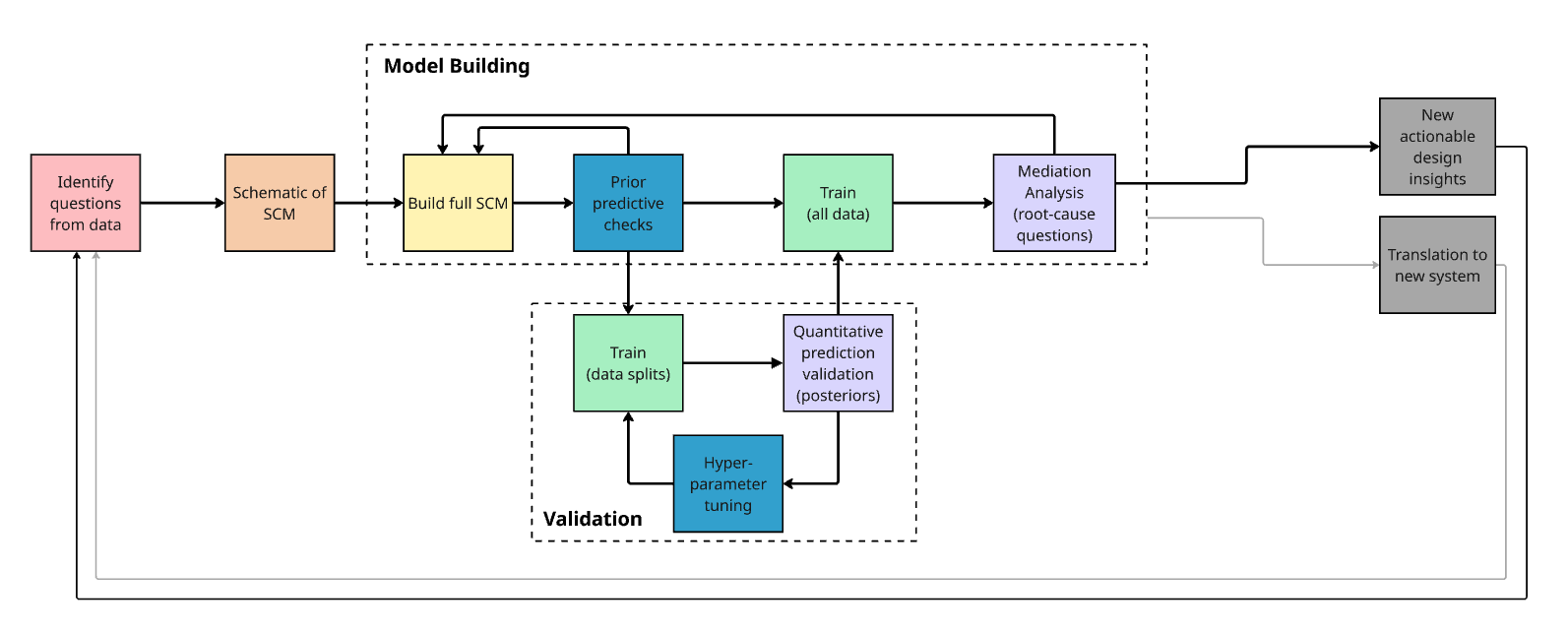}
    \caption{Causal modeling workflow summary. The workflow begins with identifying the questions with respect to an experimental dataset (or for future applications, defining an experimental dataset for the causal model from identified causal questions), then constructing the SCM, validation and training, then causal reasoning such as using mediation analyses. The final steps in the workflow are identifying actionable design insights from the model and analyses as well as translating similar models/questions/analyses to new systems. Both of the final steps loop back to the identification of questions and experimental data, with the goal of improving the translation of experimental and fundamental studies to breakthrough materials/device designs}
    \label{fig:causal-workflow}
\end{figure}

To the best of our knowledge, this work presents the first application of physics-bsed structural causal modeling to electrochemical devices. Given the complexity in modeling electrochemical systems, we present our model building workflow shown in Figure~\ref{fig:causal-workflow} as an important outcome that underscores the utility, interpretability, and credibility of the structural causal modeling framework. This workflow development was inspired by Bayesian inference workflows and causal modeling with real world data in other applications, specifically, discussions on model iteration and inclusion of domain expertise~\cite{dang2023causal, gelman2020bayesian}.

First, the system and causal questions are defined; we advocate for early inclusion of and reference to experimental datasets while constructing models in order to increase the transferability of modeling results. During the model building steps, initial model critique and evaluation is done via prior predictive checks. The model is then trained on the experimental data. Model performance in the mediation analysis is both qualitatively and quantitatively assessed to identify major modifications (as indicated by the loop back to \enquote{build full SCM} step). Separately, the model is assessed through a validation loop, whereby the train-validation-test data split is used to quantitatively assess model predictions and iteratively tune hyperparameters (kernel widths); see \nameref{methods-holdouts}. From this workflow, the root-cause analyses performed with the prediction-validated model enables deeper understanding of the experimental dataset and poses new hypotheses or actionable design insights. We present this workflow as a method for discovering new research directions in this field as well as related electrochemical systems including, but not limited to, various devices and device scales within the electrolyzer, fuel cell, and battery communities.

\section{Discussion}\label{discussion}

To summarize, we developed a structural causal model with the capability for systematic causal reasoning on how support ratio and mass loadings affect catalyst performance in an RDE experiment. This work demonstrates the first application of structural causal modeling with physics-based equations. The SCMs we developed are trained on experimental data, analogous to machine learning models. An important point to note is that the physical equations and relationships in our SCMs serve as the strong inductive bias enabling the use of small datasets (as opposed to data-driven ML methods that search for patterns without built-in constraints). To better quantify the effect of this inductive bias, we compared our SCMs' prediction performance with a collection of statistical and machine learning benchmarks; including a Bayesian linear regression~\cite{gelman1995bayesian}, a Gaussian process regression~\cite{williams2006gaussian}, and a multilayer perceptron~\cite{goodfellow2016deep}. In summary, we find porous planar SCM achieves a prediction performance at parity with the best performing baseline, while providing additional causal reasoning capabilities that the baselines do not. See Table~\ref{supp-tab:baseline-comparison} for quantitative results and additional detail on the baselines. Furthermore, the graphical representation of SCMs enables clear interpretability and explainability of the model, akin to functional block diagrams used in engineering~\cite{lind1982multilevel, rosenberg1987exploiting}. We emphasize that the SCMs used in this work are augmentations of conventional physical equations with causality and probability that enable use of causal inference tools. 

The approach we explore in this paper is similar in spirit and motivation to recent work on PINNs~\cite{raissi2019physics} and universal differential equations~\cite{rackauckas2020universal} in that both bodies of work aim to use machine learning and statistical methods to complement mechanistic models, rather than replace them. However, our work uses a SCM as the fundamental modeling substrate, rather than a collection of differential equations. Recent work has aimed to bridge these formalisms~\cite{zane2026counterfactual}; and we see both efforts as complementary.

To test the validity of the physical equations and causal assumptions used in our SCMs, we assessed the predictive performance where the porous planar SCM quantitatively out-performed the classic planar SCM, because the assumptions of a planar non-porous electrode did not hold given the experimental data. We used the quantitative metrics and the validation workflow presented to support the applicability of the porous planar SCM, outline standard operating procedures for validating SCMs, and also specifically indicate areas for model improvement. Although the porous planar SCM used in the mediation analyses can be further improved, this work is focused on the \textit{development of new methods and tools} of which model critique is included. Additional model validation and discussion is included in the SI, Figures~\ref{supp-fig:cross-validation}-~\ref{supp-fig:learning-rates}.
Briefly, the sensitivity of the porous planar model prediction and insights to the experimental dataset and prior specification was probed. Figure~\ref{supp-fig:cross-validation} shows the sensitivity of prediction accuracy to which experimental data were used for fitting using a leave-one-experiment-out cross-validation analysis. Next, Figure~\ref{supp-fig:hold-outs-mediation} shows the mediation analysis heatmap at 0.58V vs RHE for different training regimes where one of the experiments was fully omitted from the dataset; similar results to those shown in Figure 8 are observed, regardless of which experiment was left out. Similarly, in Figure~\ref{supp-fig:kernels-mediation}, varying the prior specification by changing the uncertainty distribution on various latent nodes as done in Figure~\ref{supp-fig:validation-heatmap} did not change the mediation analysis results. Finally, the overall learning rate curves for this model and dataset are shown in Figure~\ref{supp-fig:learning-rates}. Through these sensitivity tests, we demonstrate a level of robustness for the insights from this model that could both inform future experimental design and warrant experimental validation of the new hypotheses. Specific in this demonstration, the models and methods present new insights into the ORR catalyst system with entangled root causes.

We identified two key insights from the mediation analyses with the porous planar SCM. The first is the general impact of electrode thickness, whereby increasing support ratio and total material loading causes changes to the electrode thickness that can decrease catalyst performance particularly at the mass transport limiting regime. While this result is not ground-breaking~\cite{deschamps2023rotating, gridin2023gde, darab2017influence}, we disentangled and quantified the mass transport and kinetic effects at different operating potentials in the RDE experiment as well as identified regimes (the combination of support ratios, total loadings, and operating potentials) where there is trade-off between thicker electrodes and increased $N_{sites}$. The decoupled quantification is enabled by the \enquote{impossible control experiments} in the counterfactual reasoning, which requires the development and application of our SCMs. These model results highlight the need for additional experimental studies with a focus on tailoring or quantifying the electrode thickness. This could include adding electrode thickness measurements to the experimental workflow, such as atomic force microscopy on drop-cast electrodes, and or changing solvent ratios in the ink formulation to impact the morphology of drop-cast electrodes.

The second key insight from the mediation analyses is the identification of additional convolution between the support ratio and total loading on the $N_{sites}$ \underline{even when the electrode thickness is held the same}. This is suggestive of differences in utilization factors, for instance due to differences in agglomeration at different support ratios or porosity effects on utilization factors. Experiments to control the $N_{sites}$ could include changing catalyst particle size while using the same loadings and aiming to control thicknesses (catalyst active surface area and therefore $N_{sites}$ can change). The entangled variation of $N_{sites}$ with support ratio and thickness result also aligns with the highlighted areas for model improvement (shown in Figure~\ref{supp-fig:validation-heatmap}), where uncertainties remain with the utilization factors. Furthermore, there may be additional causal relationships to add between the support ratio and utilization factors. 

We  propose a few improvements to the model relevant to the insights from the causal analyses. However, we assert that such model improvements require additional characterization experiments and are beyond the scope of this work, which focuses on the development of new causal reasoning tools and methods using the classic planar and porous planar SCMs as baseline examples and first time presentations of the SCM framework. First, additional causal mechanisms may exist between support ratio and utilization factors as well as conductivity and electrode porosity and ionomer content~\cite{rao2005influence, li2021effects, su2026bridging, singh2015electrochemical}. Next, selectivity (especially for these catalysts) is an active research area, therefore modeling the selectivity needs further refinement. The activity of the Vulcan carbon support may also require additional nodes that allow future mediation analyses into understanding support optimization. Finally, careful concentration and diffusion modeling can be included; 
Figure~\ref{supp-fig:full-porous} shows an upgraded model using steady-state reaction driven diffusion total current. The same conclusions as the porous planar model are observed. Fully implementing and solving differential diffusion equations, however, requires major upgrades to our software development.
Further discussion on model improvements can be found in the SI. As the first application of SCMs to materials electrochemistry, this work has resulted in new physical SCMs and insights, as well as a proposed workflow for decoupling convoluted ink parameters of which suggested model improvements are a part.

\section{Conclusion}
Detailed understanding of the electrode as a sum of its parts is necessary to understand complex coupled phenomena as well as translate RDE experiments to predict catalyst performance in electrode assemblies and large stacked electrode devices. SCMs are a modeling framework that enables new reasoning capabilities and insights into disentangling related experimental parameters and material properties. We constructed the porous planar SCM and quantified the effect of changing support ratio and total loading on electrocatalytic performance through two different causal pathways: a electrode thickness path and a number of active sites path. We highlighted the trade off between increasing the number of active sites and increasing thickness as having opposing effects on total current density. Using the unique causal reasoning tools of counterfactuals and interventions, we also identified an additional coupled phenomena for optimizing the number of active sites through adjusting support ratio and total loading, even when constraining the electrode thickness. We apply this finding to suggest additional experiments and modeling directions. 

While the models presented in this work were constructed with a previously published experimental dataset and are specific to those experiments, we believe in the generalizability of physics-based SCMs and our proposed workflow to combine multiple published datasets and even other electrochemical systems. Therefore, while our porous-planar RDE SCM would require model updates to include data from new systems (detailed selectivity and active site modeling, ink formulation effects, etc.), the approach of using physics-based SCMs to quantitative disentangle complex pathways is a broadly applicable capability. The construction of the models, development and application of methods, and analysis of insights presented in this work required a multi-disciplinary approach, combining material science and engineering, electrocatalysis, mathematical modeling, statistics, and computer science/machine learning expertise. Furthermore, the integration of physical equations into SCMs enables the application of machine-learning-like training methods and systematic causal reasoning to small experimental datasets. Leveraging this new multi-disciplinary and accessible framework, we advocate for SCMs as a useful quantitative and visual tool in bridging material innovation and research to device performance and deployment.


\section{Methods}\label{methods}

\subsection{Experimental Details (from Kreider et al.)}\label{methods-expt}

In this work, the experimental data for training the models were sourced from published work by Kreider et al.~\cite{kreider2022strategies}. The catalyst synthesis details, electrode fabrication methods, RDE set-up, and computational details are briefly summarized here: Colloidally synthesized MnSb\textsubscript{2}O\textsubscript{6} nanoparticulate catalysts were primarily a mix of trigonal and rutile phases with small Mn\textsubscript{2}O\textsubscript{3} impurities as seen by x-ray diffraction. Mn and Sb ratios were quantified by x-ray photoelectron spectroscopy. Ex-situ surface area was measured using Brunauer-Emmett-Teller (BET) isotherm analysis of N\textsubscript{2} physisorption. Inks were made by mixing pre-determined amounts of as-synthesized MnSb\textsubscript{2}O\textsubscript{6} catalysts with commercial vulcan carbon in isopropyl alcohol and Nafion ($\approx 0.02\%$ volume). Electrodes were made drop-casting inks onto RDEs.  For electrochemical analysis the RDE used a polished exchangeable glassy carbon disc electrode (0.196 $cm^2$).  Selectivity was measured on a representative sample using a Au ring disc electrode and maximum H\textsubscript{2}O\textsubscript{2} selectivity was used for training across all potentials. All electrochemical measurements were taken in 0.1 M KOH with a Ag/AgCl reference electrode with $85 \%$ iR correction. All potentials were shifted to a reversible hydrogen electrode (RHE). CVs were done in O\textsubscript{2} saturated electrolyte at 1600 rpm. O\textsubscript{2}-saturated CVs were corrected by subtraction of CVs in N\textsubscript{2}-saturated electrolyte and only the forward or reverse scan was used for training (based on which was taken first). Double layer capacitance measurements were measured of the entire electrode at 0.5 V vs RHE from CVs at multiple scan rates in $N_2$-saturated electrolyte. DFT OH-adsorption was based on the 001 surface. 

\subsection{Structural Causal Models}

The models we explore in this paper can be described mathematically as \textbf{Structural Causal Models (SCMs)}, which has been developed within the computer science and statistics research communities over the past several decades. Informally, SCMs are defined as a collection of deterministic functions known as \textbf{structural functions} of the form $Y \leftarrow$ $f(Pa(Y), \epsilon y$), where $Pa(Y)$ represents the parents of $Y$ in a corresponding causal graphical model and $\epsilon y$ is an exogenous noise term. Note the use of an arrow, “←”, rather than an equals sign, “=”. The arrow indicates that changes in the right-hand-side of the expression cause changes in the left-hand-side, but not vice versa. In addition to defining how data is causally generated, SCMs define mathematical operations known as \textbf{interventions}, which represent how the data generating process changes as a result of some exogenous change. Conceptually, these intervention operations can be thought of as a kind of \textit{simulated controlled experiment}. For a formal mathematical definition of SCMs, interventions, and counterfactuals, see Pearl 2009~\cite{pearl2009causality}.

Causal modeling tasks are broadly grouped into two complementary categories; \textit{structure learning} and \textit{effect estimation}. Solving the structure learning problem is akin to answering the question; \enquote{given observed data on a collection of variables, which of those variables directly influence each other?} Naively searching over all possible causal structures is computationally intractable, as the number of plausible structures scales super-exponentially with the number of variables. Instead, more efficient algorithms have been developed that either (i) sequence statistical tests of conditional independence to eliminate as many candidate structures as possible~\cite{spirtes1991algorithm, spirtes2001causation, kalisch2007estimating}, or (ii) express the structure learning problem as a constrained optimization~\cite{chickering2002optimal, zheng2018dags, nazaret2025extremely} and then solve via gradient descent. Despite these computational advances, there are still fundamental limitations in our ability to learn causal structure from data. In many settings, causal structures are not identifiable~\cite{spirtes2001causation}, i.e. they can not be distinguished from even infinite observations and computation. Even in settings where causal structures are identifiable, the sheer number of candidate causal structures requires massive datasets to distinguish them. For this reason, we choose not to focus on the structure learning problem in our work.

In contrast, the effect estimation task is akin to answering the question; \enquote{given a causal structure describing which variables influence each other and a collection of observational and/or experimental data, how much does one variable influence another?} The task of effect estimation also includes variants in which we want to know the effect in a particular context or through a particular causal path. The most common approach to effect estimation involves two stages: (i) translate a causal effect estimation task into a statistical functional of observable variables, and (ii) seek out a statistical estimator of that functional with desirable statistical properties. While it outside of the scope of this paper and not directly relevant to our method, we encourage interested readers to study the do-calculus~\cite{pearl2009causality} and modern methods using efficient influence function-based estimators~\cite{kennedy2016semiparametric, agrawal2024automated}.  

A fully specified SCM enables the use of many low-level probabilistic modeling operations that can be composed to answer high-level analytical questions. What makes SCMs mathematically distinct from other probabilistic models is their support for two causal-specific mathematical operations: \textbf{intervention} and counterfactual world-splitting. Interventions represent a hypothetical (and potentially impossible) control experiment in which a variable’s value is assigned according to a user-specified value or function, and all other variables are left unchanged. This is formally modeled by producing a new SCM in which the intervened variable’s structural function $Y \leftarrow f(Pa(Y), \epsilon y)$ is replaced with the structural function $Y \leftarrow y'$, where $y'$ is the intervention assignment. Counterfactual world-splitting represents the process of imagining an alternative world, and is formally modeled by duplicating all variables and structural functions in an SCM then intervening on one or more variables. In isolation, counterfactual world-splitting provides little of value. However, when combined with interventions these counterfactual world-splitting operations can be used to formally represent rich causal questions. For example, questions about mediation analysis can be formally represented mathematically by first instantiating parallel counterfactual worlds, and then intervening on the value of mediators in one world to be the same as value in another world. For a detailed discussion, see work from Pearl~\cite{pearl2022direct}.

Furthermore, we use an extension to SCMs known as Bayesian Structural Causal Models (BSCMs)~\cite{witty2023bayesian}. Using BSCMs, modelers posit (potentially hierarchical) distributions over structural functions themselves, as well on the exogenous noise terms as in ordinary SCMs. In this way, “learning” structural causal models from data is reduced to joint probabilistic inference as in ordinary Bayesian statistical methods a la Bayes rule, and often approximated using approximation algorithms. This provides a principled framework for representing broad uncertainty, to reason about how that uncertainty is updated in light of data, and how that uncertainty propagates to the causal questions that are often of interest in scientific domains.

\subsubsection{Translating experimental workflows and physical equations into SCMs}\label{methods-building SCMs}

In this work, we translated conventional physical equations into SCMs, whereby the structure (e.g. edges connecting nodes) encodes causality. Specifically, the directionality of the edges in the causal graphical model representations (e.g., Figure~\ref{fig:SCM-schematic}c) stipulate a cause going to an effect. This directionality necessitates that changing a “cause” will impact an “effect”. For instance, the Butler-Volmer equation is translated into a causal relationship in which an applied overpotential \textit{causes} a change in the total current density. The directionality provides the interpretability of the physics in the SCM during model simulations and data fitting as well as enabling interventions which can be composed to answer causal queries. Specifically, for the RDE experiments modeled in this work, overpotential is the experimental control being directly applied, therefore we construct this directed causal relation. This can be a different approach to modeling and fitting compared to conventional equation fitting: the Butler-Volmer equation (Tafel approximation) is often rearranged so that $\eta = (slope) \times log_{10}(i/i_{0})$, however, in this rearrangement, current density or exchange current density do not cause overpotential. For the conversion relationships used in this model (e.g., catalyst mass [mg]/geometric area [cm\textsuperscript{2}] $\rightarrow$ catalyst mass loading [mg$*$cm\textsuperscript{-2}]), we ascribed a causal direction to match an \enquote{experimental workflow}: a target total material loading and support ratio is selected by the experimentalist $\rightarrow$ individual amounts of catalyst and support are calculated (and weighed out) $\rightarrow$ the catalyst and support loadings are calculated from the electrode geometric area. We note that causality in these conversion relationships may appear contrived, however, future work aims to address the equivalency of these conversion relationships. The causal directions specified, especially in equivalent conversion relationships dictate the specific causal questions we can answer related to those variables.

Another structural element in the SCM is the hierarchical organization of repeated structures in the causal model into \textbf{plates}. Essentially, this captures the multiple individual experiments and measurements performed in an experimental dataset, where the support ratios and the total material mass loadings were changed, and for each experiment, a cyclic voltammogram was measured. While the support ratios and the total mass loadings in the different experiments may differ, the causal relationships between mass, surface area, thickness, etc. are the same across the experiments (these are repeated structures). The first plate in our SCM is the experiment plate to represent all experiments across a set. Similarly, while the applied potential is different for each data point in the cyclic voltammetry measurement, which consists of many potential-current data points, the way applied potential affects overpotential, kinetic current density, etc. remains the same across the timepoints. This hierarchical plating structure is graphically represented by the boxes the framing nodes in Figure~\ref{fig:SCM-schematic}.

The third important element of our physics-based SCMs to note is specification of uncertainty. As mentioned, the mathematical structural functions in SCMs include exogenous noise terms. In our physics-based materials-device SCMs, we apply the noise terms as probability distributions that represent uncertainty stemming from measurement errors and uncertainty in the mechanisms or values (epistemic uncertainty). An example of uncertainty in the measurement is the Gaussian distribution we applied to the total current node to represent instrument error ($\sim$0.01 mA/cm\textsuperscript{2}). An example of uncertainty in the values may be seen in the porosity node; porosity is challenging to measure/not measured in the experimental dataset, therefore we give the porosity root node an initial value of 0.5 with a beta probability “kernel”. The \textbf{kernel} is a functional which accepts parameters as inputs and returns distributions. This is effectively stating that our initial belief about porosity, is some value centered around 0.5, but may be slightly larger or smaller. This also means that if the model is used to run many simulations, with each simulation sampling from the porosity prior distribution, all downstream nodes will necessarily also be a distribution rather than single scalars. An example of uncertainty in the mechanisms can be demonstrated with the total ECSA node; total ECSA is caused by the ECSA\textsubscript{cat} and ECSA\textsubscript{sup}: if the catalyst had an ECSA of 20 cm\textsuperscript{2} while the support had an ECSA of 15 cm\textsuperscript{2}, then the total electrode would have an ECSA of 35 cm\textsuperscript{2}. This value would then be combined with a Gaussian kernel to generate a Gaussian distribution centered at  35 cm\textsuperscript{2}. This distribution would then be used to represent both the value and uncertainty of our estimates for ECSA. Note that the applied Gaussian distribution to the Total ECSA is added in addition to the distribution of values caused by the incoming nodes, which are also distributions. Additionally, specific to the prior generation, we applied uniform kernels to several of the nodes that we intended to sweep. For instance, the applied voltage node was ascribed a uniform kernel centered at 0.6 V with lower bound $\approx$0 V, and upper bound of 1.2 V. This allowed the generation of the voltammogram prior distributions. 

\subsubsection{Mediation Analysis}\label{methods-mediation}

Conventional sensitivity analyses perturb model inputs and measure the resulting output changes~\cite{saltelli2008global}.
For example, varying \gls{m} while monitoring \gls{j} gives the finite-difference sensitivity $\Delta j / \Delta m$.
Similarly, varying the \gls{phi} gives $\Delta j / \Delta \phi$.
Specifically, let model $F$ calculate \gls{j} from \gls{m} and \gls{phi}:

\begin{equation}
j \coloneqq F(m,\phi)
\end{equation}

\noindent
Then assign base-case values $m := m_0$ and $\phi := \phi _0$ to calculate $j_0$:

\begin{equation}
j_0 \coloneqq F(m_0,\phi_0)
\end{equation}

\noindent
and perturbing \gls{m} to $m_1$ while holding \gls{phi} at $\phi _0$ gives the sensitivity $S_m$:

\begin{align}
j_m &\coloneqq F(m_1,\phi_0) \\
S_m &\coloneqq \frac{j_m-j_0}{m_1-m_0}
\end{align}

\noindent
This sensitivity is affected by the choice of units though, so it is difficult to compare to other sensitivity values.
Thus we also normalize by the standard deviations to yield a unitless sensitivity \gls{Sm}:

\begin{align}
S^{\sigma}_m &\coloneqq \frac{\sigma_m}{\sigma_j} \cdot \frac{j_m-j_0}{m_1-m_0}
\end{align}

\noindent
where $\sigma_m$ and $\sigma_j$ are the standard deviations of the electrode loading data and current density data, respectively.
Analogously, perturbing \gls{phi} to $\phi_1$ while holding \gls{m} at $m_0$ gives \gls{Sphi}:

\begin{align}
j_\phi &\coloneqq F(m_0,\phi_1) \\
S_\phi &\coloneqq \frac{j_\phi-j_0}{\phi_1-\phi_0} \\
S^{\sigma}_\phi &\coloneqq \frac{\sigma_{\phi}}{\sigma_j} \cdot \frac{j_\phi-j_0}{\phi_1-\phi_0}
\end{align}

\noindent
Thus the higher a sensitivity, the more important the corresponding input.

Such sensitivities are useful for understanding systems and models, but there are fewer existing methods for follow-up questions such as \enquote{What portion of \gls{Sm} comes from changes in \gls{h} vs. changes in \gls{nsites}?}
Causal mediation analyses seek to answer such questions.
More broadly, causal reasoning is not an alternative to already-existing sensitivity analysis methods, but rather a supplemental tool that helps extract even more insight.

Specifically, causal mediation analyses isolate pathway effects by simulating control experiments in which changing inputs such as \gls{m} affects intermediate variables such as \gls{h}, while interventions lock other intermediate variables such as \gls{nsites} to reference values as if \gls{m} was never changed.
Thus when effects are propagated to model outputs such as \gls{j}, only responses from \gls{h} are observed, but not responses from \gls{nsites}.
The result of this procedure is formally defined as the natural direct effect (NDE)~\cite{pearl2022direct}, providing a quantitative metric on the effect of a variable on another through a particular causal path.

This is shown in Figures~\ref{fig:causal-pathway},~\ref{fig:mediation}, \&~\ref{fig:heatmaps} where we quantify how simultaneous changes in \gls{m} and \gls{phi} affect \gls{j} through \gls{h} without any \gls{nsites}-mediated effects.
These figures also show the converse where---e.g., how \gls{m} and \gls{phi} affect \gls{j} through \gls{nsites} without any \gls{h}-mediated effects.

We provide a simplified illustration of the NDE calculations for our RDE example as an illustration below, omitting a description of sampling from probability kernels for ease of exposition.
Specifically, let one part of the SCM $g$ calculate \gls{h} and \gls{nsites} from \gls{m} and \gls{phi}, and another part of our SCM $f$ calculate \gls{j} from the \gls{h} and \gls{nsites}:

\begin{align}
(h,N) &\coloneqq g(m,\phi) \\
j &\coloneqq f(h,N)
\end{align}

\noindent
where \gls{nsites} is shortened to $N$ for brevity.
We then specify two references cases $0$ and $1$ perturbing the total mass loading $m$ and support:catalyst ratio $\phi$:

\begin{align}
(h_0,N_0) &\coloneqq g(m_0,\phi_0) \\
j_{00} &\coloneqq f(h_0,N_0) \\
(h_1,N_1) &\coloneqq g(m_1,\phi_1) \\
j_{11} &\coloneqq f(h_1,N_1)
\end{align}

\noindent
Next we do a simulated control experiment where we perturb $m := m_1$ and $\phi := \phi_1$, but intervene on $N := N_0$ rather than allowing its natural value $N_1$.
This intervention yields $j_{10}$:

\begin{align}
(h_1,N_1) &\coloneqq g(m_1,\phi_1) \\
N &\coloneqq N_0 \\
j_{10} &\coloneqq f(h_1,N_0)
\end{align}

\noindent
An analogous intervention fixes $h$ instead:

\begin{align}
(h_1,N_1) &\coloneqq g(m_1,\phi_1) \\
h &\coloneqq h_0 \\
j_{01} &\coloneqq f(h_0,N_1)
\end{align}

\noindent
Lastly, the natural direct effects~\cite{pearl2009causality} from thickness $NDE_h$ and natural direct effects from number of active sites $NDE_N$ are calculated by referencing the intervened values against the base reference value:

\begin{align}
NDE_{h} &\coloneqq j_{10}-j_{00} \\
NDE_{N} &\coloneqq j_{01}-j_{00}
\end{align}

In summary, conventional sensitivity analyses calculate changes in outputs such \acrlong{j} relative to one-at-a-time changes in directly controllable variables like \acrlong{m} or \acrlong{phi}.
These analyses are useful for understanding how inputs affect outputs.
When we seek to understand more deeply \textit{why} these inputs affect the outputs, causal mediation analyses enable us to quantify pathway-mediated effects between inputs and outputs.
Note that these pathway-specific effects may not sum to the total effects---e.g., $j_{11} - j_{00} \ne NDE_h + NDE_N$.

\subsubsection{Causal Probabilistic Programming}\label{methods-causalprobprog}

Causal probabilistic programs automate many of the symbolic and numerical operations used in probabilistic and causal modeling and reasoning, including interventions, counterfactual world-splitting, probabilistic sampling, probabilistic density function evaluation, and many more. These systems build on existing software systems for differentiable tensor computations, which have been developed extensively for machine learning workflows. We use ChiRho~\cite{chiro-repo}, a causal probabilistic programming language built on the Pyro probabilistic programming language~\cite{bingham2019pyro} and the PyTorch tensor computation system~\cite{paszke2019pytorch}. These systems automatically translate causal models and causal questions into numerical operations that themselves can be automated using automatic differentiation~\cite{paszke2017automatic} and compiled to massively parallel GPU hardware. See Figure~\ref{supp-fig:flowchart} for a schematic diagram of causal probabilistic programming systems.

\subsection{Model Training}\label{training}

We used stochastic variational inference (SVI) to approximate the posterior distributions over SCM model parameters, which propagate to uncertainty estimates over causal effect estimates~\cite{hoffman2013stochastic}. SVI algorithms involve first defining a variational family over a collection of random variables X and parameterized by $\phi$, denoted $Q(X;\phi)$. For example, $Q$ could be a collection of Gaussian random variables where $\phi$ represents the mean and variance of each Gaussian distribution. To approximate a posterior distribution, SVI involves optimizing $\phi$ to minimize an upper bound on the Kullbeck-Leibler divergence between $Q$ and the (unknown) posterior distribution $P(X|Data)$, a quantity known as the evidence lower-bound (ELBO). Even though the true posterior distribution $P(X|Data)$ is unknown, SVI algorithms can still optimize $\phi$ by computing stochastic estimates of the gradient of the ELBO from available numerical quantities, and then using those stochastic gradients in an iterative gradient descent algorithm. This procedure effectively reduces probabilistic inference, which involves solving an intractable integral, into a numerical optimization procedure that can be scaled to hundreds of thousands of parameters and data points.

The kernels described earlier also serve as ways to add domain knowledge and subject matter expertise to SCMs and can affect model training. For example, if we are certain that the total electrode loading is a sum of the support loading and catalyst loading, then the kernel we apply to total electrode loading should have a narrow standard deviation. This leads to narrow prior distributions for total electrode loading. Narrow kernels incentivize model training to stay near the initial guesses, while wide kernels allow model training to diverge from initial guesses, where “initial guesses” for variables comprise both the kernel and the structural equation for the variable. This idea of implicit incentivization is referred to as “regularization” in optimization and machine learning communities. Thus the kernels we choose are model regularizers.

We used a clipped Adam optimizer, with a 5$\times$10\textsuperscript{-4} learning rate, no learning rate decay, and 4,000 training steps.

\subsubsection{Data hold-outs}\label{methods-holdouts}

Different data hold-out strategies were utilized during the modeling process. For the initial modeling building, all data used for training and evaluated on mediation results. For fine-tuning and model refinement (adjusting initial default values, uncertainty parameters), the model was trained on the training set (n = 6), then quantitatively assessed on the validation set (n = 2). After this iterative refinement, the model was trained on the combined training and validation set (n = 8) and quantitatively assessed on the test set (n = 1) for predictive performance. Lastly, the causal mediation analysis for new insights utilized the model trained on the entire dataset. 

\subsection{Validation}\label{validation}

The metrics used for quantitative model assessment include (1) the accuracy of the predictions, (2) the goodness-of-fit, which is whether the model’s errors show leftover patterns~\cite{kutner1984applied}, and (3) the quality of the predictions’ uncertainties~\cite{tran2020methods}. We assess model prediction accuracy using Mean Absolute Error (MAE), Root Mean Squared Error (RMSE), and coefficient of determination (R\textsuperscript{2}). MAE weights all errors equally and is common in the machine learning community, while RMSE disproportionately punishes outliers and is common in the statistics community. R\textsuperscript{2} measures how much of the spread in the data can be explained by the model’s predictions. It ranges from 0–1, where 0 corresponds to a model that simply guesses the average and 1 corresponds to a perfect model.

Model goodness-of-fit (i.e., whether the model’s residuals have patterns) was assessed using Pearson’s correlation coefficient (r), residual diagnostics, and posterior predictive checks. Pearson’s correlation coefficient (r) measures whether the model’s residuals follow a straight-line pattern when predictions are plotted against observations; it ranges from -1 to 1, with 1 corresponding to a perfect linear correlation. 

Model uncertainty was quantified using Expected Calibration Error (ECE). The probability distributions from our models are used to create prediction intervals and assessed on their confidence. For example, a calibrated 80$\%$ prediction interval (10th-90th percentile) contains 80$\%$ of the data. ECE measures the difference between our prediction intervals and reality. A higher ECE indicates the model reports more confidence despite being less accurate and is shown mathematically below:
\begin{center}
    $ECE = \sum_{m=1}^{M}w_{m}\left| p_{m}-\widehat{p}_{m} \right|$
\end{center}
where $M$ = number of prediction interval levels we evaluate (we used 0.01, 0.02, … 0.99); $p_{m}$ = the width of the prediction interval level; $\widehat{p}_{m}$ = the fraction of observed values that actually fell within the model's $p_{m}$ prediction interval; and $w_{m}$ = weight for interval $m$ (we used a uniform weight of 1 for all values of $m$). 

We also assess model accuracy and uncertainty simultaneously using Expected Log Predictive Density (ELPD). ELPD evaluates where the actual data point falls on the predicted distribution and records the probability the model assigned there. High probability at the true value indicates the model anticipated the outcome well; low probability indicates it was surprised. ELPD combines these probabilities across all data points and rewards models that consistently place high probability on what actually occurred. High values of ELPD are better — e.g., -50 is better than -200.

\subsection{Glossary of terms}\label{glossary}

\begin{itemize}

\item \textbf{Structural Causal Model}: mathematical modeling framework which includes cause-and-effect relationships between variables. The relationships are modeled by structural functions. SCMs can be represented in a causal graph, specifically a directed acyclic graph in which there are no loops formed by nodes and edges.

\item \textbf{Node}: the variables in the structural causal models. Root nodes have no upstream causes.

\item \textbf{Edge}: the causal relationship between nodes. Graphically represented by directional arrows connecting nodes

\item \textbf{Mechanism} or \textbf{Structural Function}: mathematical functions that describe a node as a function of its direct causes. An example for the structural function $f(a,b) \rightarrow c$ could be $a \times b \rightarrow c$. 

\item \textbf{Kernel}: functionals that intake parameter values and return distributions. We specify and use several types of distributions such as normal, folded normal, beta, etc.

\item \textbf{Plate}: structural elements of the SCM that denote repeating independent units, such as different experimental samples. Graphically represented as boundary boxes that group the repeated nodes. 

\item \textbf{Prior distribution}: simulated distributions (for the nodes in the model) before training the model on data.

\item \textbf{Training}: the model is conditioned on data and approximate posterior distributions are updated in order to better reflect the data. In this work, stochastic variational inference was used during model training. 

\item \textbf{Inference}: In the context of Bayesian modeling, inference is the process of estimating approximate posterior distributions conditioned on data. This should not be confused with inference in the broader machine learning and neural network research communities, where inference is the calculation of graph outputs given graph inputs.

\item \textbf{Posterior distribution}: simulated distributions after the training

\item \textbf{Intervention}: deliberate constraining or setting a variable to (e.g., different from its natural value). Also known as a "do-operator" in the causal research community.

\item \textbf{Counterfactual reasoning}: comparing \enquote{what if} interventions. In the SCM framework, evaluating the different outcomes when different node values are artificially constrained/set.

\end{itemize}

\section{Acknowledgments}

Our paper used previous data from Dr. Melissa Kreider.  We appreciate that her work was easily accessible and formatted in a manner that was easy to use. We also appreciate Dr. Kreider's thoughtful comments that helped us refine this work. We thank our co-workers at TRI for insightful conversations, specifically Dr. Steven Torrisi, Dr. Rumen Iliev, Dr. Joseph Montoya, and Dr. Brian Storey. 

\section{Author contributions}
E.W., L.H., S.W., M.B.S., and K.T. jointly conceived the project, developed the structural causal model, interpreted the results, and wrote and edited the manuscript. E.W. led the writing efforts. S.W. and K.T co-led the causal modeling software and computational development, and M.B.S. and E.W. co-led the electrocatalysis experimental and material physics integration efforts.

\section{Competing Interests}
The authors declare no competing interests.

\section{Data availability} 
Source data are provided with this paper. See GitHub repositories below.

\section{Code availability}\label{code}
The code used to perform this work is available at:
\texttt{https://github.com/TRI-AMDD/Whyzen-Public} and \texttt{https://github.com/TRI-AMDD/Whyzen-RDE-Public}

 
\putbib
\end{bibunit}

\clearpage
\appendix

\clearpage
\appendix

\renewcommand{\thetable}{S\arabic{table}}
\renewcommand{\thefigure}{S\arabic{figure}}
\setcounter{table}{0}
\setcounter{figure}{0}

\begin{bibunit}

\setcounter{page}{1}

\begin{titlepage}
    \thispagestyle{empty}
    \centering

    \vspace*{0.18\textheight}

    {\LARGE\bfseries Supporting Information\par}

    \vspace{1.5cm}

    {\Large
    Bridging electrode preparation and electrocatalyst performance
    with physics-based causal AI using Bayesian Structural Causal Models\par}

    \vspace{1.5cm}

    {\large
    Evelyna Wang, Linda Hung, Samuel Witty,\\
    Michaela Burke Stevens, and Kevin Tran\par}

    \vfill
\end{titlepage}

\newcolumntype{L}[1]{>{\RaggedRight\arraybackslash}p{#1}}

\newpage

\begin{landscape}

\scriptsize

\setlength{\LTcapwidth}{\linewidth}
\setlength{\tabcolsep}{3pt}
\renewcommand{\arraystretch}{1.5}

\begin{longtable}{
|L{0.12\linewidth}
|L{0.08\linewidth}
|L{0.08\linewidth}
|L{0.11\linewidth}
|L{0.08\linewidth}
|L{0.17\linewidth}
|L{0.08\linewidth}
|L{0.12\linewidth}
|L{0.12\linewidth}|
}

\caption{Full list of nodes in the Classic Planar SCM. We categorize the node types as data from Kreider et al. or other literature, calculated, and fitted. Nodes can also be categorized as electrode, device, catalyst, or support parameters, or electrocatalytic performance properties. The mechanism for each node (or initial values for root nodes) are also listed. The last columns list the kernel type applied (kernels intake a value and output a distribution) as well as the rationale for whether/why the kernel was applied.}
\label{supp-tab:classic-planar-nodes}\\

\hline
\normalsize\textbf{Node name} &
\normalsize\textbf{Symbol} &
\multicolumn{3}{L{0.27\linewidth}|}{\normalsize\textbf{Type}} &
\normalsize\textbf{Mechanism or Initial Value} &
\normalsize\textbf{Units} &
\normalsize\textbf{Kernel} &
\normalsize\textbf{Rationale} \\
\hline
\endfirsthead

\hline
\normalsize\textbf{Node name} &
\normalsize\textbf{Symbol} &
\multicolumn{3}{L{0.27\linewidth}|}{\normalsize\textbf{Type}} &
\normalsize\textbf{Mechanism or Initial Value} &
\normalsize\textbf{Units} &
\normalsize\textbf{Kernel} &
\normalsize\textbf{Rationale} \\
\hline
\endhead

Total Loading &
$\mathrm{Area\ load}_{\mathrm{tot}}$ &
Data &
Electrode properties &
Root node &
Range $[0.09,1.25]$ varies by experiment &
mg cm\textsuperscript{-2} &
-- &
Conditioned on data \\
\hline

Ratio Support:Catalyst &
-- &
Data &
Electrode properties &
Root node &
Range $(0,1)$ varies by experiment &
unitless $(0,1)$ &
-- &
Conditioned on data \\
\hline

Geometric area &
$A_{\mathrm{cell}}$ &
Data &
Electrode properties &
Root node &
$0.196$ &
cm$^{-2}$ &
-- &
Conditioned on data \\
\hline

Temperature &
$T$ &
Data &
Device settings &
Root node &
$298$ &
K &
-- &
Conditioned on data \\
\hline

Rotation Rate &
-- &
Data &
Device settings &
Root node &
$1600$ &
radians &
-- &
Conditioned on data \\
\hline

Catalyst Composition &
$\mathrm{comp}_{\mathrm{cat}}$ &
Data &
Catalyst properties &
Root node &
MnSb$_2$O$_6$ &
N/A &
-- &
Conditioned on data \\
\hline

Support Mass &
$\mathrm{Mass}_{\mathrm{sup}}$ &
Data &
Support properties &
-- &
$\mathrm{Ratio}_{\mathrm{sup}}\times \mathrm{Total\ loading}\times A_{\mathrm{cell}}$ &
mg &
-- &
Conditioned on data \\
\hline

Catalyst Mass &
$\mathrm{Mass}_{\mathrm{cat}}$ &
Data &
Catalyst properties &
-- &
$(1-\mathrm{Ratio}_{\mathrm{sup}})\times \mathrm{Total\ loading}\times A_{\mathrm{cell}}$ &
mg &
-- &
Conditioned on data \\
\hline

Normalized Support Loading &
$\mathrm{Area\ load}_{\mathrm{sup}}$ &
Data &
Support properties &
-- &
$\mathrm{Mass\ load}_{\mathrm{sup}}/A_{\mathrm{cell}}$ &
mg cm$^{-2}$ &
-- &
Conditioned on data \\
\hline

Normalized Catalyst Loading &
$\mathrm{Area\ load}_{\mathrm{cat}}$ &
Data &
Catalyst properties &
-- &
$\mathrm{Mass\ load}_{\mathrm{cat}}/A_{\mathrm{cell}}$ &
mg cm$^{-2}$ &
-- &
Conditioned on data \\
\hline

Applied Potential &
$V$ &
Data &
Electrocatalytic performance &
Root node &
$[0,1.2]$ &
V &
-- &
Conditioned on data \\
\hline

Reference Potential &
$V_{\mathrm{ref}}$ &
Data &
Electrocatalytic performance &
Root node &
$1.23$ &
V &
-- &
Conditioned on data \\
\hline

Applied overpotential &
$\eta$ &
Calculated &
Electrocatalytic performance &
-- &
$-(V - V_{\mathrm{ref}})$ &
V &
None &
Deterministic; if applied potential and reference voltage are known, applied overpotential can be directly calculated. \\
\hline

Specific Surface Area Support &
$\mathrm{SSA}_{\mathrm{sup}}$ &
Data \cite{kreider2022strategies} &
Support properties &
Root node &
$3854$ &
cm$^2$ g$^{-1}$ &
-- &
Conditioned on data \\
\hline

Specific Surface Area Catalyst &
$\mathrm{SSA}_{\mathrm{cat}}$ &
Data \cite{kreider2022strategies} &
Catalyst properties &
Root node &
$60$ &
cm$^2$ g$^{-1}$ &
-- &
Conditioned on data \\
\hline

Support Volumetric Density &
$\rho_{\mathrm{sup}}$ &
Data \cite{cabot_vulcan_xc72} &
Support properties &
Root node &
$1700$ &
mg cm$^{-3}$ &
-- &
Conditioned on data \\
\hline

Catalyst Volumetric Density &
$\rho_{\mathrm{cat}}$ &
Data \cite{nalbandyan2015new} &
Catalyst properties &
Root node &
$6000$ &
mg cm$^{-3}$ &
-- &
Conditioned on data \\
\hline

Specific Capacitance Support &
$C_{\mathrm{sup}}$ &
Data \cite{kreider2022strategies} &
Support properties &
Root node &
$0.0045$ &
mF cm$^{-2}$ &
-- &
Conditioned on data \\
\hline

Specific Capacitance Catalyst &
$C_{\mathrm{cat}}$ &
Data \cite{kreider2022strategies} &
Catalyst properties &
Root node &
$0.051$ &
mF cm$^{-2}$ &
-- &
Conditioned on data \\
\hline

Molecular Weight &
$MW_{\mathrm{cat}}$ &
Fitted &
Catalyst properties &
-- &
$\sum_i N_{i,\mathrm{at}}\mathrm{wt}_i$ &
g mol$^{-1}$ &
Folded Relative Normal, rel. std. $=0.01$ &
Small uncertainty to represent deviations in molecular weight due to surface oxidation or impurities. \\
\hline

Site Areal Density &
$\rho_{\mathrm{site}}$ &
Calculated &
Catalyst properties &
-- &
$\frac{1}{\mathrm{SSA}_{\mathrm{cat}}\times MW}$ &
mol cm$^{-2}$ &
None &
Deterministic; any distribution is due to uncertainty in parent nodes. \\
\hline

Adsorbate Binding Energy &
$\Delta G_{\mathrm{HO}^*}$ &
Data \cite{kreider2022strategies} &
Catalyst properties &
Root node &
$0.92$ &
eV &
-- &
Conditioned on data \\
\hline

Turn Over Frequency &
$\mathrm{ToF}$ &
Fitted \cite{zhang2024unraveling} &
Catalyst properties &
-- &
$f(\Delta G_{\mathrm{HO}^*})$ &
s$^{-1}$ &
Folded relative normal, rel. std. $=0.05$ &
Uncertainty in approximation from volcano plots. \\
\hline

Support Surface Area &
$\mathrm{SA}_{\mathrm{sup}}$ &
Calculated &
Support properties &
-- &
$\mathrm{SSA}_{\mathrm{sup}}\times \mathrm{Mass\ Load}_{\mathrm{sup}}$ &
cm$^2$ &
None &
Deterministic; any distribution is due to uncertainty in parent nodes. \\
\hline

Catalyst Surface Area &
$\mathrm{SA}_{\mathrm{cat}}$ &
Calculated &
Catalyst properties &
-- &
$\mathrm{SSA}_{\mathrm{cat}}\times \mathrm{Mass\ Load}_{\mathrm{cat}}$ &
cm$^2$ &
None &
Deterministic; any distribution is due to uncertainty in parent nodes. \\
\hline

Support Thickness &
$\delta_{sup}$ &
Calculated &
Support properties &
-- &
$\rho_{\mathrm{sup}}\times \mathrm{Area\ load}_{\mathrm{sup}}$ &
cm &
None &
Deterministic; any distribution is due to uncertainty in parent nodes. \\
\hline

Catalyst Thickness &
$\delta_{cat}$ &
Calculated &
Catalyst properties &
-- &
$\rho_{\mathrm{cat}}\times \mathrm{Area\ load}_{\mathrm{cat}}$ &
cm &
None &
Deterministic; any distribution is due to uncertainty in parent nodes. \\
\hline

Support Utilization &
$U_{\mathrm{echem,sup}}$ &
Fitted &
Support properties &
Root node &
$0.5$ &
unitless $[0,1]$ &
Beta, var. $=0.1$ &
Difficult to measure; treated as fit parameter; beta distributions are bounded $[0,1]$. \\
\hline

Catalyst Utilization &
$U_{\mathrm{echem,cat}}$ &
Fitted &
Catalyst properties &
Root node &
$0.5$ &
unitless $[0,1]$ &
Beta, var. $=0.1$ &
Difficult to measure; treated as fit parameter; beta distributions are bounded $[0,1]$. \\
\hline

Electrochemical Surface Area Support &
$\mathrm{ECSA}_{\mathrm{sup}}$ &
Calculated &
Support properties &
-- &
$\mathrm{SA}_{\mathrm{sup}}\times U_{\mathrm{echem,sup}}$ &
cm$^2$ &
None &
Deterministic; any distribution is due to uncertainty in parent nodes. \\
\hline

Electrochemical Surface Area Catalyst &
$\mathrm{ECSA}_{\mathrm{cat}}$ &
Calculated &
Catalyst properties &
-- &
$\mathrm{SA}_{\mathrm{cat}}\times U_{\mathrm{echem,cat}}$ &
cm$^2$ &
None &
Deterministic; any distribution is due to uncertainty in parent nodes. \\
\hline

Double Layer Capacitance Support &
$C_{\mathrm{DL,sup}}$ &
Calculated &
Support properties &
-- &
$C_{\mathrm{sup}}\times \mathrm{ECSA}_{\mathrm{sup}}$ &
mF &
None &
Deterministic; any distribution is due to uncertainty in parent nodes. \\
\hline

Double Layer Capacitance Catalyst &
$C_{\mathrm{DL,cat}}$ &
Calculated &
Catalyst properties &
-- &
$C_{\mathrm{cat}}\times \mathrm{ECSA}_{\mathrm{cat}}$ &
mF &
None &
Deterministic; any distribution is due to uncertainty in parent nodes. \\
\hline

Electrolyte Diffusivity &
$D_{\mathrm{electrolyte}}$ &
Data &
Device settings &
Root node &
$1.9\times10^{-5}$ &
cm$^2$ s$^{-1}$ &
-- &
Conditioned on data \\
\hline

Reactant bulk concentration &
$C_b$ &
Data &
Device settings &
Root node &
$1.2\times10^{-6}$ &
mol cm$^{-3}$ &
-- &
Conditioned on data \\
\hline

Electrolyte kinematic viscosity &
$\nu$ &
Data &
Device settings &
Root node &
$0.01$ &
m$^2$ s$^{-1}$ &
-- &
Conditioned on data \\
\hline

Total Electrochemical Surface Area &
$\mathrm{ECSA}_{\mathrm{active}}$ &
Fitted &
Electrode properties &
-- &
$\mathrm{ECSA}_{\mathrm{sup}}+\mathrm{ECSA}_{\mathrm{cat}}$ &
cm$^2$ &
Folded relative normal, rel. std. $=10^{-1}$ &
Uncertainty in direct summation of $\mathrm{ECSA}_{\mathrm{cat}}$ and $\mathrm{ECSA}_{\mathrm{sup}}$. \\
\hline

Surface Area electrode &
$\mathrm{SA}_{\mathrm{elec}}$ &
Fitted &
Electrode properties &
-- &
$\mathrm{SA}_{\mathrm{sup}}+\mathrm{SA}_{\mathrm{cat}}$ &
-- &
Folded relative normal, rel. std. $=10^{-5}$ &
Uncertainty in direct summation of $\mathrm{SA}_{\mathrm{cat}}$ and $\mathrm{SA}_{\mathrm{sup}}$. \\
\hline

Total Thickness &
$\delta_{\mathrm{tot}}$ &
Fitted &
Electrode properties &
-- &
$\delta_{\mathrm{sup}}+\delta_{\mathrm{cat}}$ &
cm &
Folded relative normal, rel. std. $=0.05$ &
Uncertainty in direct summation of catalyst and support thicknesses. \\
\hline

Porosity &
$\epsilon$ &
Fitted &
Electrode properties &
Root node &
$0.5$ &
unitless $[0,1]$ &
Beta, var. $=0.05$ &
Difficult to measure; treated as fit parameter; beta distributions are bounded $[0,1]$. \\
\hline

Electrode Diffusivity (pores) &
$D_{\mathrm{electrode}}$ &
Calculated &
Electrode properties &
-- &
$D_{\mathrm{electrolyte}}\epsilon^{3/2}$ &
cm$^2$ s$^{-1}$ &
None &
Deterministic; any distribution is due to uncertainty in parent nodes. \\
\hline

Fraction Active Sites &
$f_{\mathrm{active}}$ &
Fitted &
Electrode properties &
Root node &
$10^{-4}$ &
unitless $[0,1]$ &
Log normal, std. $=0.05$, base $=10$ &
Difficult to measure; treated as fit parameter; log normal applied. \\
\hline

Total Double Layer Capacitance &
$C_{\mathrm{DL}}$ &
Data &
Electrode properties &
-- &
$C_{\mathrm{DL,sup}}+C_{\mathrm{DL,cat}}$ &
mF &
-- &
Conditioned on data \\
\hline

Number of Sites &
$N_{\mathrm{sites}}$ &
Calculated &
Electrode parameter &
-- &
$f_{\mathrm{active}}\times \mathrm{ECSA}_{\mathrm{active}}\times \rho_{\mathrm{site}}$ &
mol &
None &
Deterministic; any distribution is due to uncertainty in parent nodes. \\
\hline

Alpha &
$\alpha$ &
Fitted &
Electrocatalytic performance &
Root node &
$0.5$ &
unitless $[0,1]$ &
Beta, var. $=10^{-3}$ &
Difficult to measure; treated as fit parameter; log normal applied. \\
\hline

Exchange Current Density &
$j_0$ &
Fitted &
Electrocatalytic performance &
-- &
$\frac{N_{\mathrm{sites}}\times \mathrm{ToF}\times F\times n}{A_{\mathrm{cell}}}$ &
mA cm$^{-2}$ &
Folded normal, std. $=10^{-4}$ &
Uncertainty in calculation using turnover frequency rather than reactant concentrations. \\
\hline

Selectivity &
$n$ &
Data [Kreider et al.] &
Electrocatalytic performance &
Root node &
Range $[2.6,3.96]$ varies by experiment &
mol &
-- &
Conditioned on data \\
\hline

Kinetic Current Density &
$j_k$ &
Calculated &
Electrocatalytic performance &
-- &
$j_0\exp\left(-\frac{\alpha\times F\times n_\times \eta}{RT}\right)$ &
mA cm$^{-2}$ &
None &
Deterministic; any distribution is due to uncertainty in parent nodes; testing mechanism validity. \\
\hline

Limiting Current Density &
$j_{\mathrm{lim}}$ &
Calculated &
Electrocatalytic performance &
-- &
$0.62nFD^{2/3}\nu^{-1/6}C_b\omega^{1/2}$ &
mA cm$^{-2}$ &
None &
Deterministic; any distribution is due to uncertainty in parent nodes; testing mechanism validity. \\
\hline

Total Current Density &
$j$ &
Data [Kreider et al.] &
Electrocatalytic performance &
-- &
$\left(\frac{1}{j_k}+\frac{1}{j_{\mathrm{lim}}}\right)^{-1}$ &
mA cm$^{-2}$ &
Normal, std. $=10^{-2}$ &
Conditioned on data, with small uncertainty for instrument error. \\
\hline

\end{longtable}
\end{landscape}

\normalsize

\newpage
\begin{figure}[htbp]
    \centering
    \includegraphics[width=\textwidth]{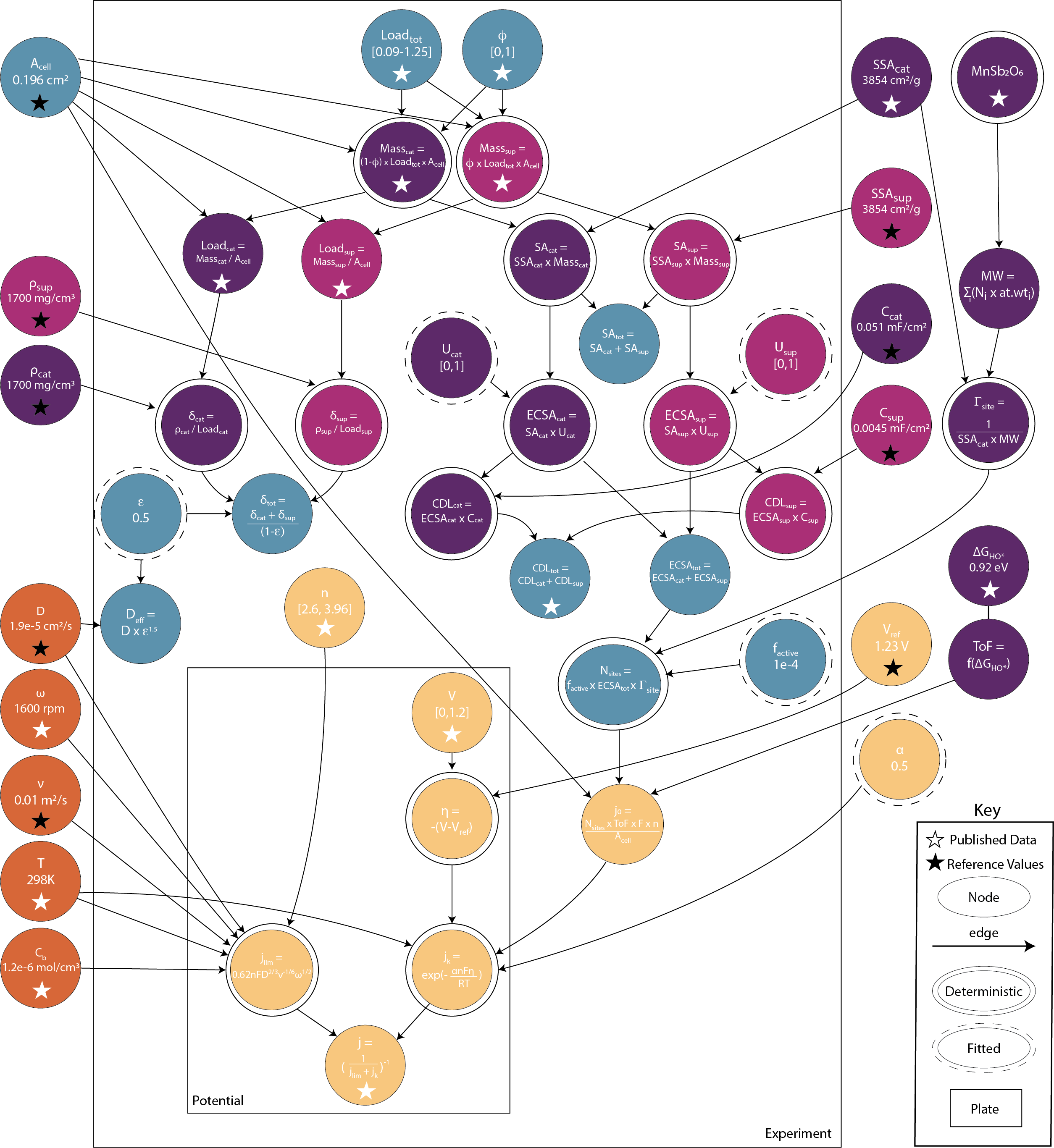}
    \caption{The Classic Planar SCM (as shown in the main text Figure~\ref{fig:SCM-schematic}) shown with the mechanisms listed in the model. Compare to Table~\ref{supp-tab:classic-planar-nodes}}
    \label{supp-fig:classic-planar-SCM}
\end{figure}

\newpage
\begin{figure}[htbp]
    \centering
    \includegraphics[width=\textwidth]{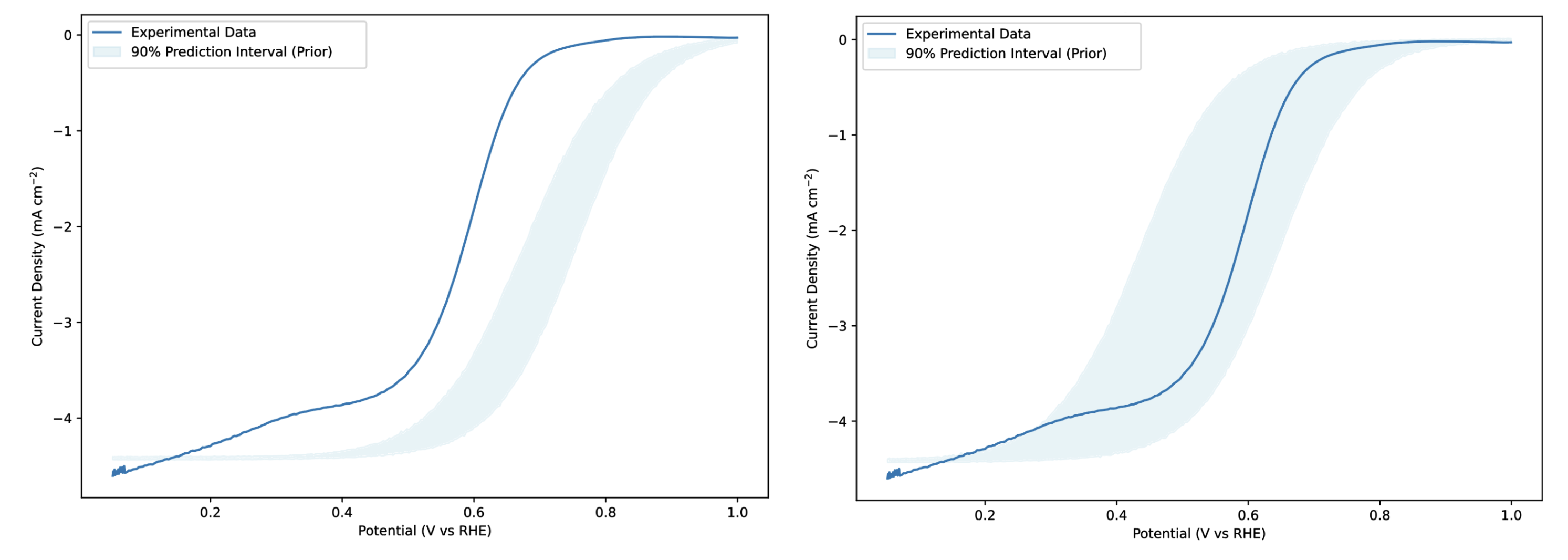}
    \caption{Prior predictive checks showing an example of a poor prior distribution (left) and an acceptable prior on the (right) as compared to experiment A. These checks are done before training on the data.}
    \label{supp-fig:priors-predictive}
\end{figure}

\newpage
\begin{figure}[htbp]
    \centering
    \includegraphics[width=\textwidth]{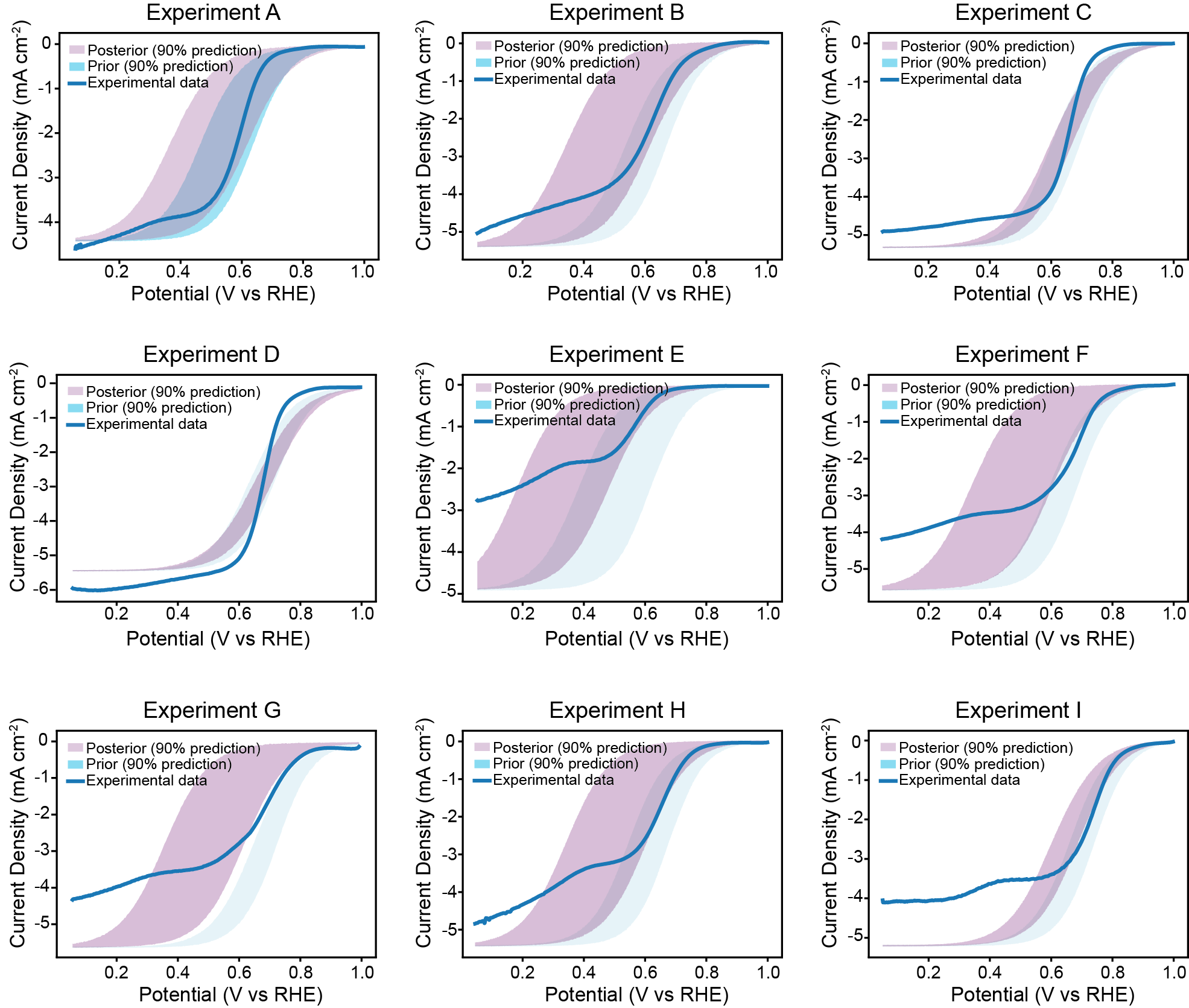}
    \caption{Voltammogram priors and posteriors from the classic planar SCM for all experiments. Posterior predictions were generated from the classic planar SCM trained on all experiments.}
    \label{supp-fig:classic-planar-voltammograms}
\end{figure}

\newpage
\begin{figure}[htbp]
    \centering
    \includegraphics[width=\textwidth]{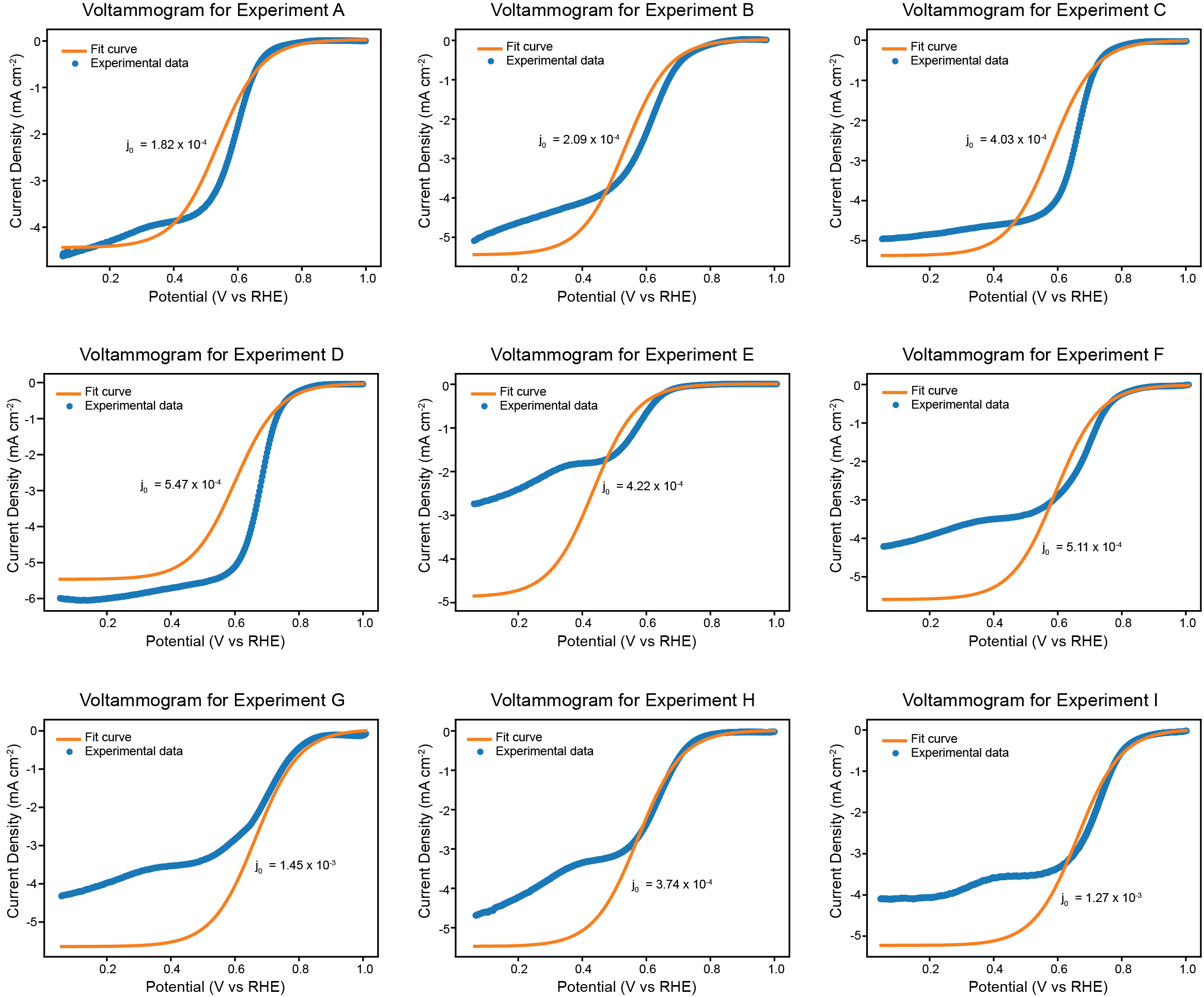}
    \caption{Voltammogram plots from fittings to the Levich, Butler-Volmer, Koutecky-Levich equations. The n selectivity values from Kreider et al. were used to help with the fitting \cite{kreider2022strategies}. A single global charge transfer coefficient was used during the fit across all experiments ($\alpha$ = 0.375).}
    \label{supp-fig:levich-fitting}
\end{figure}

\newpage
\begin{figure}[htbp]
    \centering
    \includegraphics[width=\textwidth]{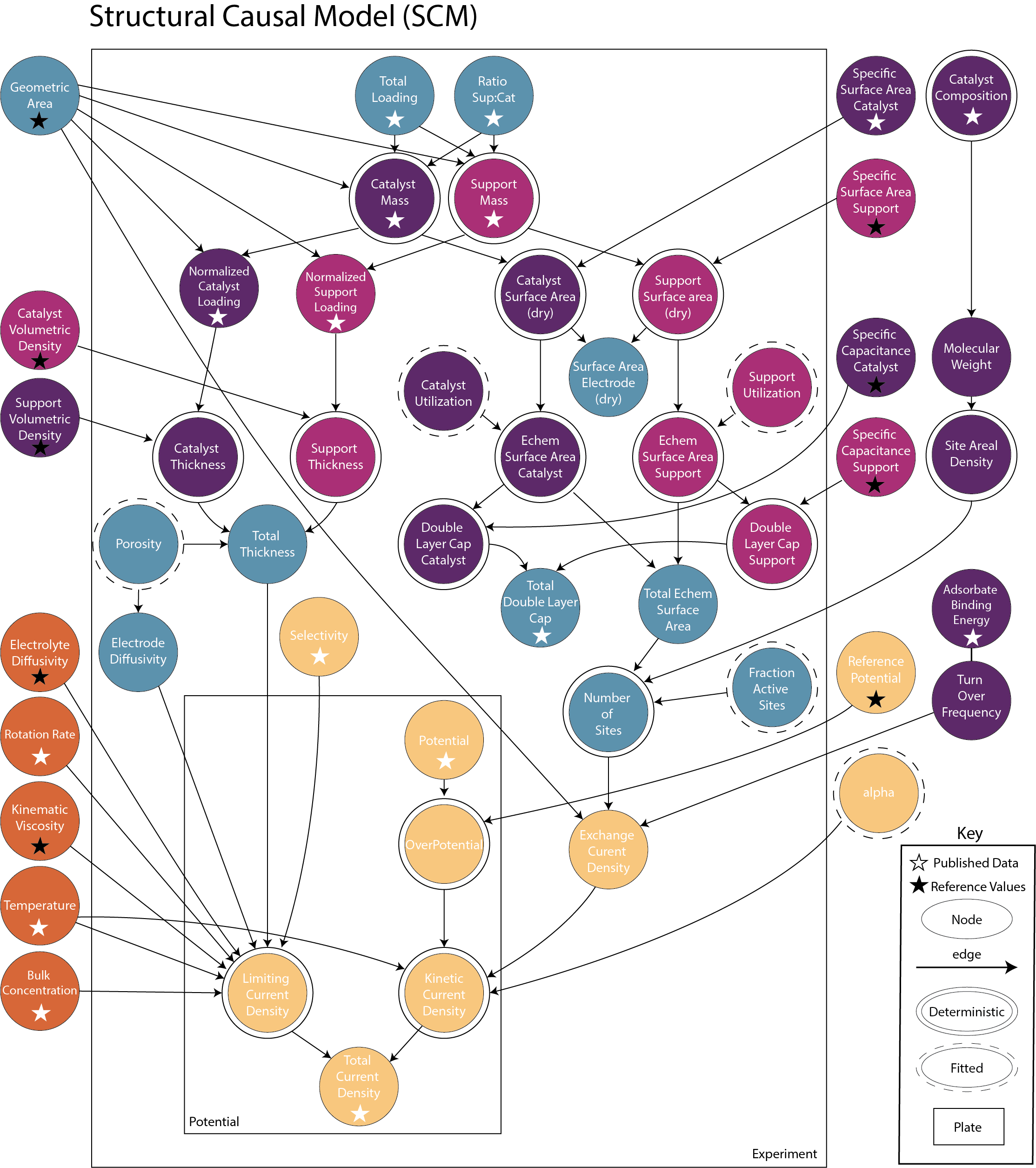}
    \caption{Full view of the graphical representation for the porous planar SCM, showing the added edges connecting thickness and diffusivity in the porous electrode to the limiting current density node.}
    \label{supp-fig:scm-porous}
\end{figure}

\newpage
\begin{figure}[htbp]
    \centering
    \includegraphics[width=\textwidth]{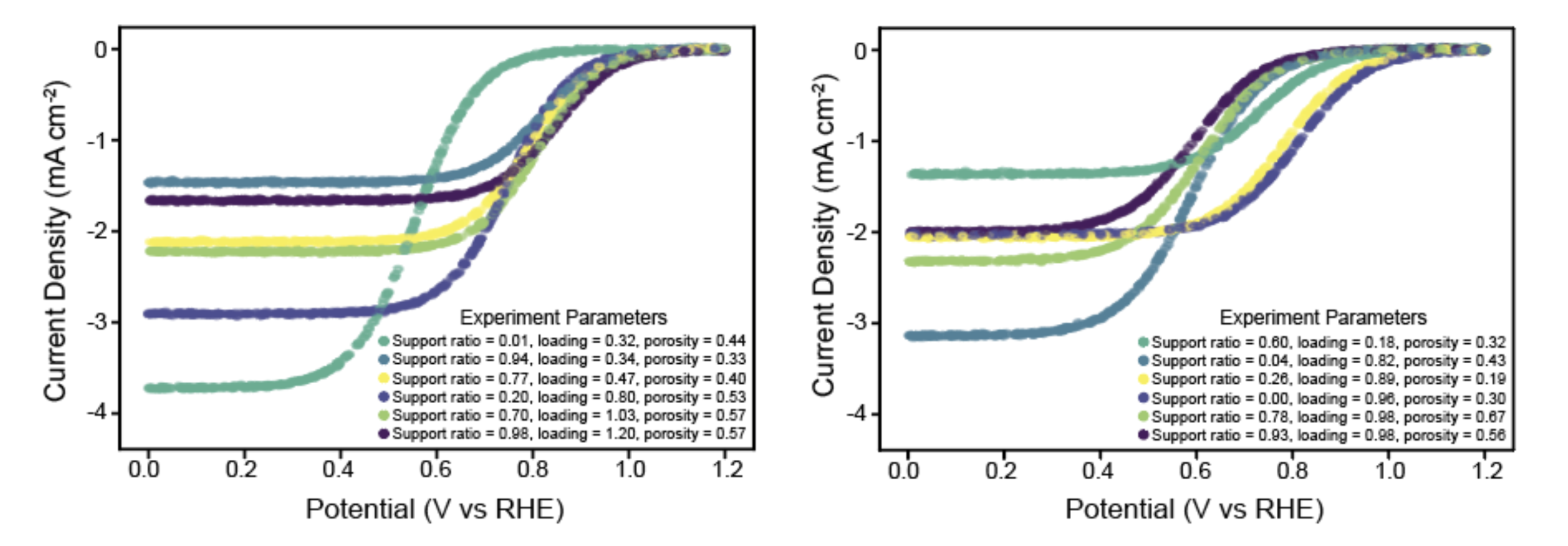}
    \caption{Simulated voltammogram priors for the porous planar SCM. Each voltage trace represents a different simulated experiment (with different samples for the support ratio, total loading, porosities, utilization factors, fraction active sites, etc.). The curves within a plot share the same sampled global values (e.g., charge transfer coefficient).}
    \label{supp-fig:porous-priors}
\end{figure}

\newpage
\begin{figure}[htbp]
    \centering
    \includegraphics[width=\textwidth]{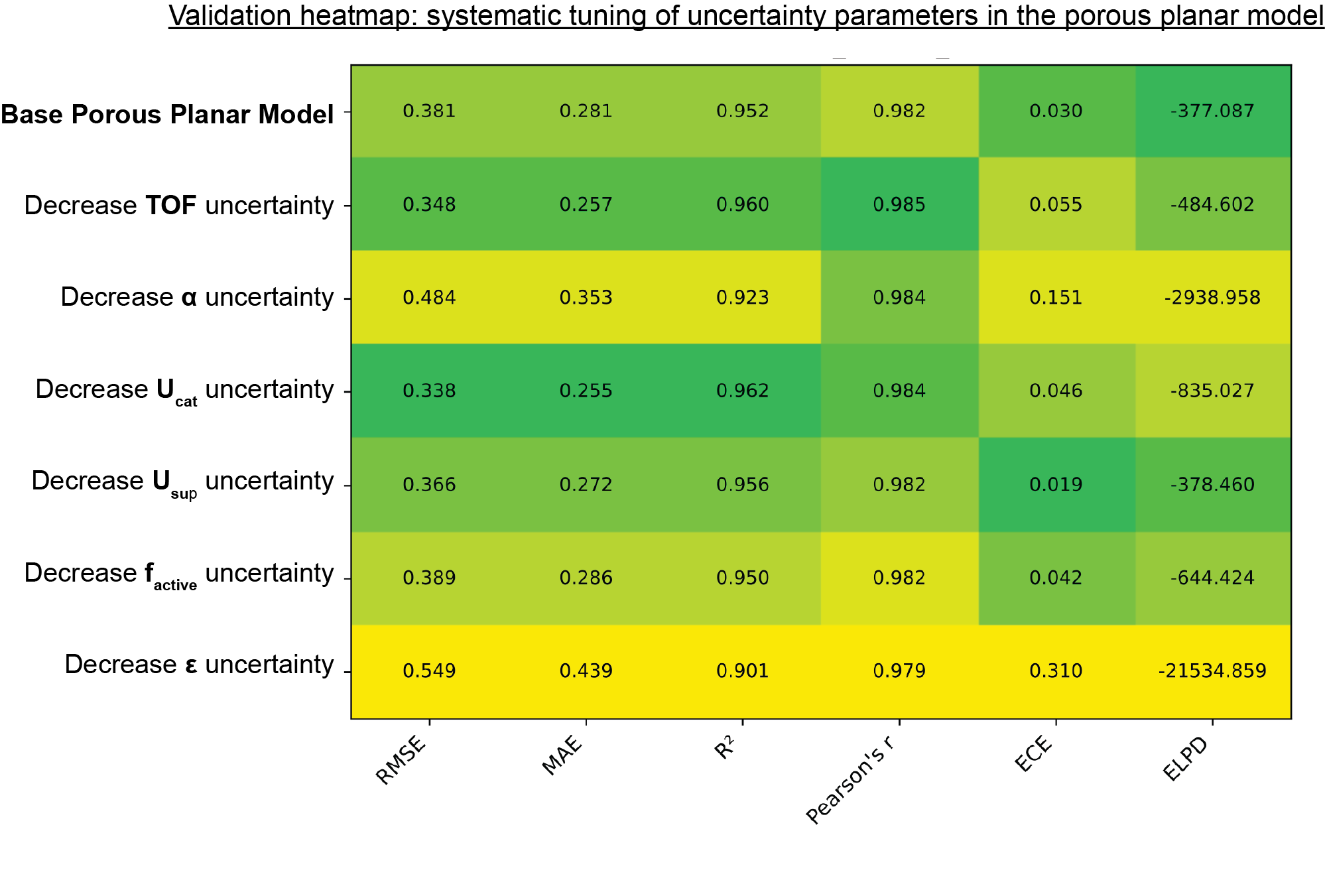}
    \caption{Validation heatmap for the porous planar SCM varying the kernel widths applied at each node for the turnover frequency, charge transfer coefficient, catalyst utilization factor, support utilization factor, fraction of active sites, and porosity. The color map was added by ranking each column for ‘best’ to ‘worst’ value then coloring from green to yellow. This assessment was performed with the training-validation data split: the uncertainty parameters for each of these nodes were separately and systematically adjusted, then the model performance was validated (on the validation data set). We observe that decreasing the allowed uncertainty for the catalyst utilization and porosity led to the largest decrease in residual analysis of predictive performance. Decreasing the kernel width for the fraction of active sites also decreased the accuracy and uncertainty evaluated from the ELPD metric (despite a slight improvement in the residuals).}
    \label{supp-fig:validation-heatmap}
\end{figure}

\newpage
\begin{landscape}

\begingroup
\setlength{\tabcolsep}{10pt}
\renewcommand{\arraystretch}{1.5}

\begin{table}[!htbp]
\centering

\caption{Sensitivity analysis of the porous planar model prediction performance with respect to sweeps of parameter values. Underlined values indicate default parameter settings. Bold values indicate the best $R^2$ and RMSE within each parameter sweep.}
\label{table:sensitivity-analysis}

\begin{footnotesize}
\begin{tabular}{|p{3.1cm}|c|c|c|}
\hline
\textbf{Parameter} &
\textbf{Sweep values (default underlined)} &
\textbf{$R^2$ (best value in bold)} &
\textbf{RMSE (best value in bold)} \\
\hline
\hline

\multirow[t]{5}{3.1cm}{Turnover frequency [s$^{-1}$]}
& $10^{-4}$ & 0.933 & 0.451 \\ 
& $10^{-3}$ & 0.901 & 0.535 \\ 
& \underline{$6.8 \times 10^{-3}$} & \textbf{0.953} & \textbf{0.380} \\
& 0.1 & 0.578 & 1.13 \\
& 1 & 0.461 & 1.28 \\
\hline

\multirow[t]{5}{3.1cm}{Charge transfer coefficient [unitless]}
& 0.3 & 0.898 & 0.577 \\ 
& \underline{0.4} & \textbf{0.953} & \textbf{0.380} \\ 
& 0.5 & 0.738 & 0.893 \\ 
& 0.7 & 0.223 & 1.54 \\ 
& 0.95 & -0.455 & 2.11 \\
\hline

\multirow[t]{5}{3.1cm}{Catalyst utilization factor [unitless]}
& 0.1 & \textbf{0.964} & \textbf{0.331} \\ 
& 0.3 & 0.949 & 0.395 \\ 
& \underline{0.5} & 0.953 & 0.380 \\
& 0.7 & 0.955 & 0.370 \\ 
& 0.95 & 0.953 & 0.379 \\
\hline

\multirow[t]{5}{3.1cm}{Support utilization factor [unitless]}
& 0.2 & 0.925 & 0.479 \\
& 0.35 & 0.935 & 0.445 \\ 
& \underline{0.5} & \textbf{0.953} & \textbf{0.380} \\ 
& 0.7 & 0.873 & 0.621 \\ 
& 0.95 & 0.852 & 0.672 \\
\hline

\multirow[t]{5}{3.1cm}{Fraction of active sites [unitless]}
& $10^{-5}$ & 0.909 & 0.526 \\ 
& $3 \times 10^{-5}$ & 0.926 & 0.475 \\ 
& \underline{$10^{-4}$} & \textbf{0.953} & \textbf{0.380} \\ 
& $10^{-3}$ & 0.745 & 0.882 \\ 
& $10^{-2}$ & 0.800 & 0.780 \\
\hline

\multirow[t]{5}{3.1cm}{Electrode porosity [unitless]}
& 0.3 & \textbf{0.966} & \textbf{0.320} \\ 
& 0.35 & 0.959 & 0.353 \\ 
& \underline{0.5} & 0.953 & 0.380 \\ 
& 0.65 & 0.946 & 0.406 \\ 
& 0.8 & 0.952 & 0.382 \\
\hline

\end{tabular}
\end{footnotesize}

\end{table}
\endgroup

\end{landscape}

\newpage
\begin{figure}[htbp]
    \centering
    \includegraphics[width=\textwidth]{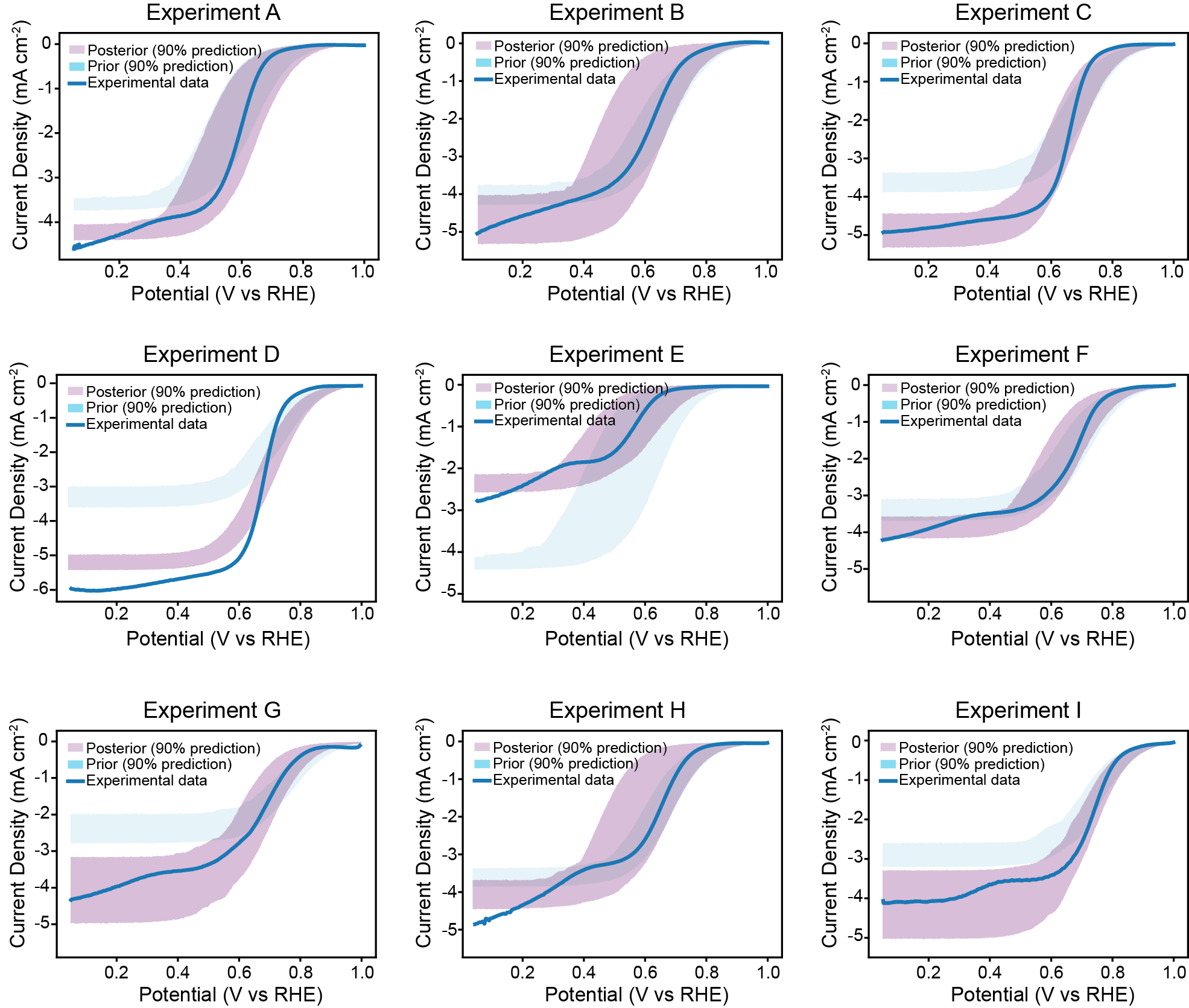}
    \caption{Voltammogram priors and posteriors from the porous planar SCM for all experiments. Posterior predictions were generated from the porous planar SCM trained on all experiments. We observed a better fit to the mass transport limited regime as compared to the predictions shown in Figure S3 from the classic planar model.}
    \label{supp-fig:porous-planar-voltammograms}
\end{figure}

\newpage
\begin{figure}[htbp]
    \centering
    \includegraphics[width=\textwidth]{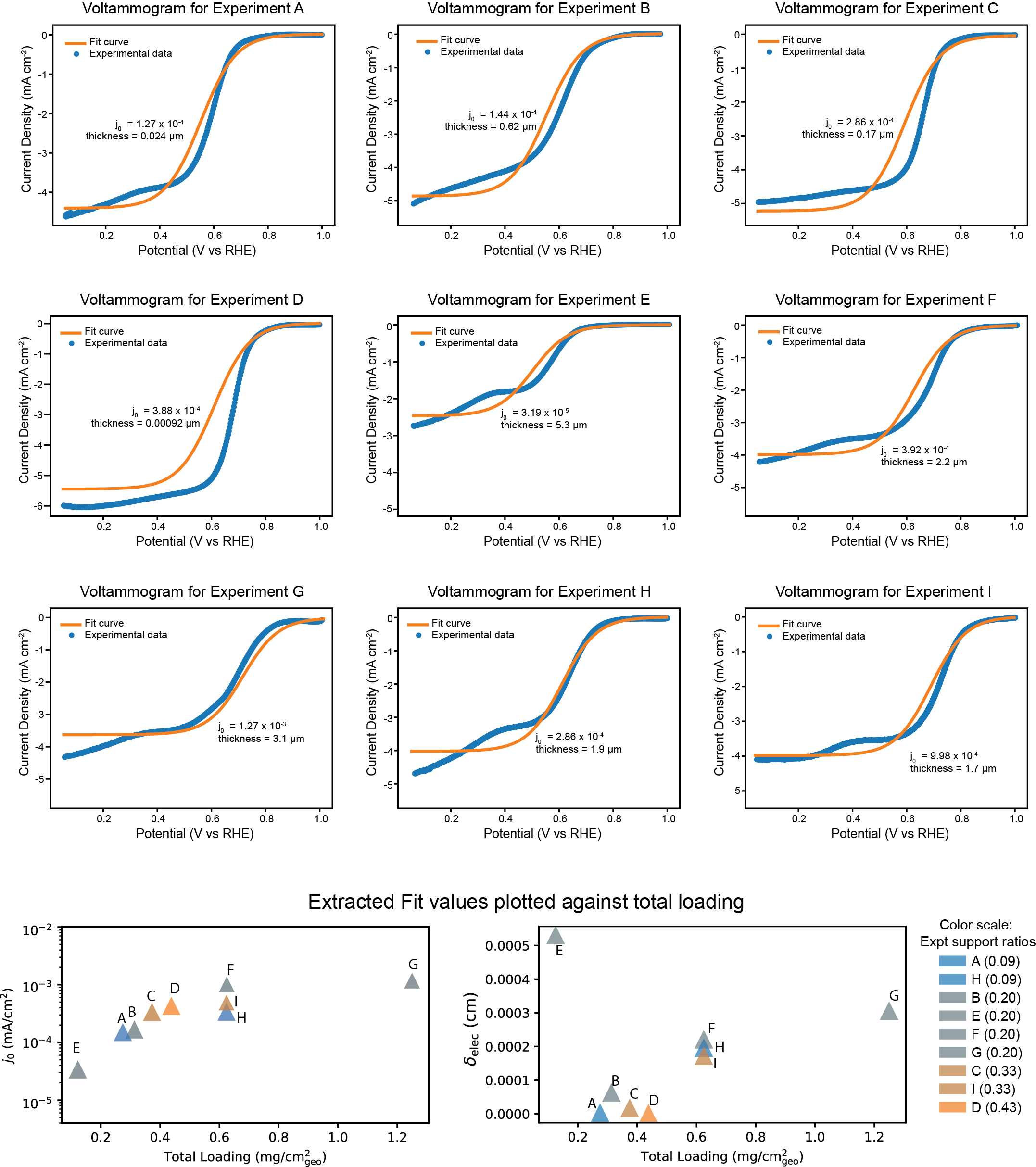}
    \caption{Voltammogram plots from fittings to the Modified Levich, Butler-Volmer, Koutecky-Levich equations. The n selectivity calculated from Kreider et al. were used to help with the fitting. A single global charge transfer coefficient was used during the fit across all experiments ( = 0.396). The extracted fit values for exchange current density and thickness are also shown as a function of total loading.}
    \label{supp-fig:modified-levich-fitting}
\end{figure}

\newpage
\begin{figure}[htbp]
    \centering
    \includegraphics[width=\textwidth]{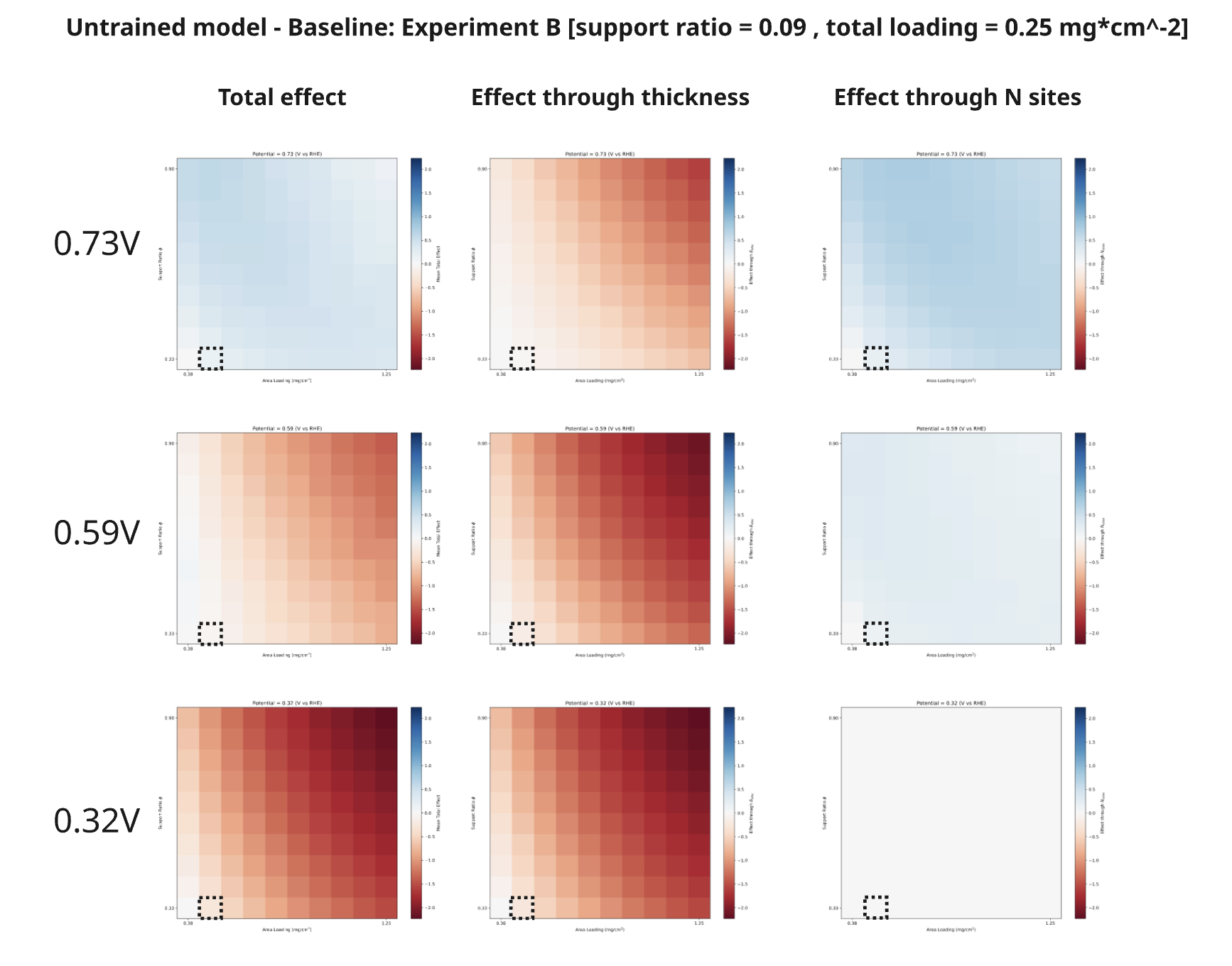}
    \caption{Mediation analyses priors (before training on experimental data) shown with respect to the baseline simulation [support ratio = 0.10 and total loading = 0.27 mg*cm$^{-2}$]}
    \label{supp-fig:mediation-heatmaps-untrained}
\end{figure}

\clearpage
\begin{figure}[htbp]
    \centering
    \includegraphics[width=\textwidth]{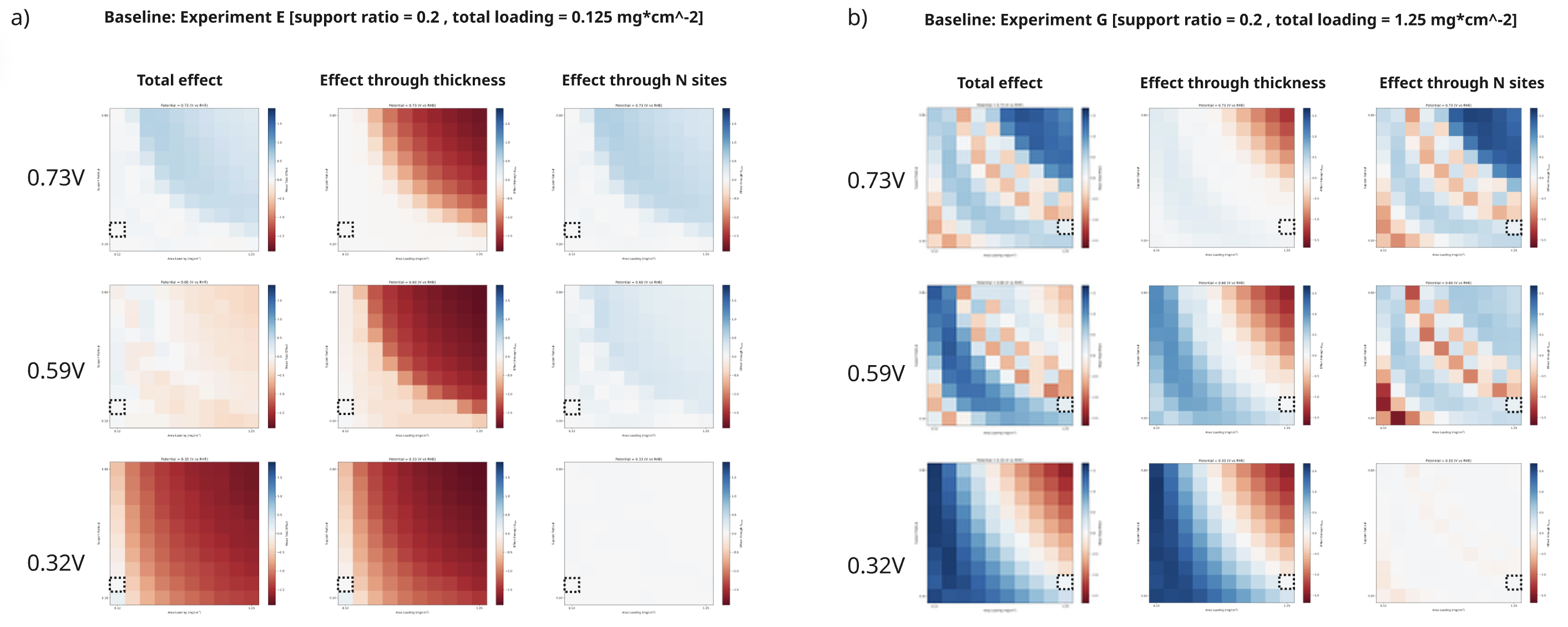}
    \caption{a) the simulations are compared against experiment E [support ratio = 0.2, total loading = 0.125 mg*cm$^{-2}$]. Similar trends to Figure 8 main text are observed. At moderate loadings and support ratios, there is some optimum in increasing current density until the mass transport regime, where the effect of increased thickness dominates in decreasing the current density. b) The simulations are compared against experiment G [support ratio = 0.2, total loading = 1.25 mg*cm$^{-2}$]. Again similar trends are observed, however from a slight opposite perspective. Starting at the highest loading, a decrease in total loading can improve the total current density. Decreasing both the total loading and increasing the support ratio can lead to a decrease in the N sites (going diagonally up to the left from experiment G), thereby decreasing the total current density.}
    \label{supp-fig:mediation-heatmaps-others}
\end{figure}

\newpage
\begin{figure}[htbp]
    \centering
    \includegraphics[width=\textwidth]{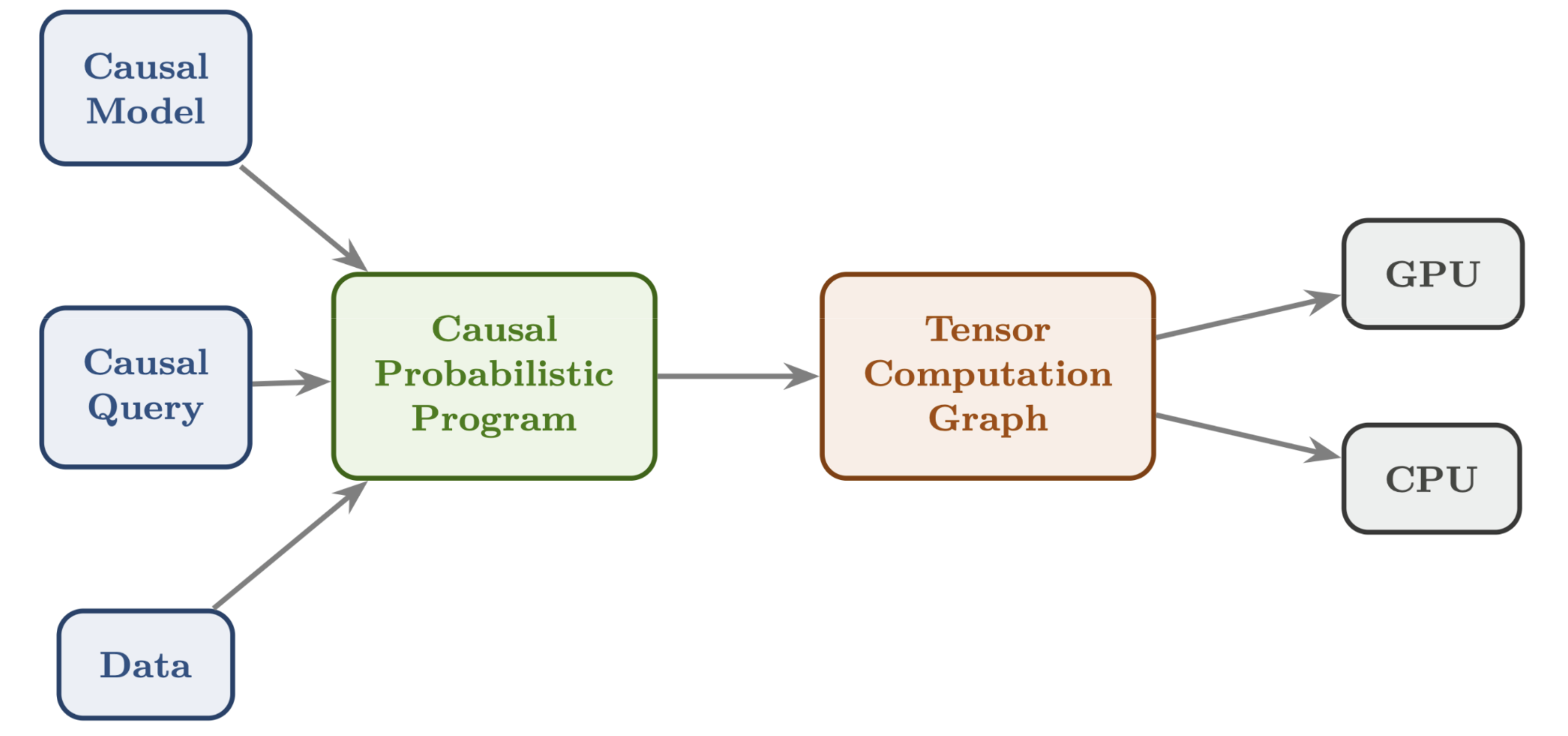}
    \caption{Schematic representation of a causal modeling programming system, which intakes the causal model, queries, and experimental data and applies symbolic and numerical operations in the form of tensor computations. These rely on small relatively small amounts of GPU, especially compared to large ML models}
    \label{supp-fig:flowchart}
\end{figure}

\newpage
\begin{landscape}

\begin{table}[p]
    \centering
\caption[
    Baseline comparison on held-out experiment H
]{
    Comparison of current-density predictions on held-out experiment H. All non-mechanistic baselines were trained on the raw, uninterpolated current observations from the training and validation electrodes and use the same observed non-target inputs available to the RDE models. The nonlinear GP uses a Mat\'ern-$3/2$ kernel with a separate learned length scale for each varying input. Both the GP and the Bayesian linear regression use a fixed Gaussian observation-noise standard deviation of $0.01~\mathrm{mA\,cm^{-2}}$. The MLP is a single network with two 32-unit hidden layers, $\tanh$ activations, and 10\% dropout. Its uncertainty is estimated from 500 Monte Carlo dropout passes combined with the same fixed observation noise. GP ELPD values use the analytic Gaussian predictive density, whereas MLP ELPD is estimated by applying log-mean-exp to the fixed Normal likelihood evaluated at the latent dropout predictions. The mechanistic SCM results are those reported for experiment H in Figure~4. Arrows indicate the preferred metric direction, and boldface identifies the best reported value in each column.
}
    \label{supp-tab:baseline-comparison}

    \begin{tabular}{llrrrrrr}
        \toprule
        Model & Type
        & RMSE $\downarrow$
        & MAE $\downarrow$
        & $R^2$ $\uparrow$
        & Pearson's $r$ $\uparrow$
        & ECE $\downarrow$
        & ELPD $\uparrow$ \\
        \midrule

        Bayesian linear regression
        & Non-mechanistic
        & 3.255
        & 3.228
        & $-2.478$
        & 0.9711
        & 0.500
        & $-4.83 \times 10^{7}$ \\

        Matérn GP
        & Non-mechanistic
        & 0.900
        & 0.731
        & 0.734
        & 0.9882
        & 0.191
        & $\mathbf{-2{,}158}$ \\

        MLP (32, 32)
        & Non-mechanistic
        & 0.382
        & 0.285
        & $\mathbf{0.952}$
        & $\mathbf{0.9932}$
        & 0.067
        & $-13{,}655$ \\

        Classic planar SCM
        & Mechanistic
        & 0.94
        & 0.73
        & 0.71
        & 0.98
        & 0.36
        & $-2.2 \times 10^{7}$ \\

        Porous planar SCM
        & Mechanistic
        & $\mathbf{0.38}$
        & $\mathbf{0.27}$
        & 0.95
        & 0.98
        & $\mathbf{0.02}$
        & $-1.0 \times 10^{4}$ \\
        \bottomrule
    \end{tabular}
\end{table}
\end{landscape}

\newpage
\begin{figure}[htbp]
    \centering
    \includegraphics[width=\textwidth]{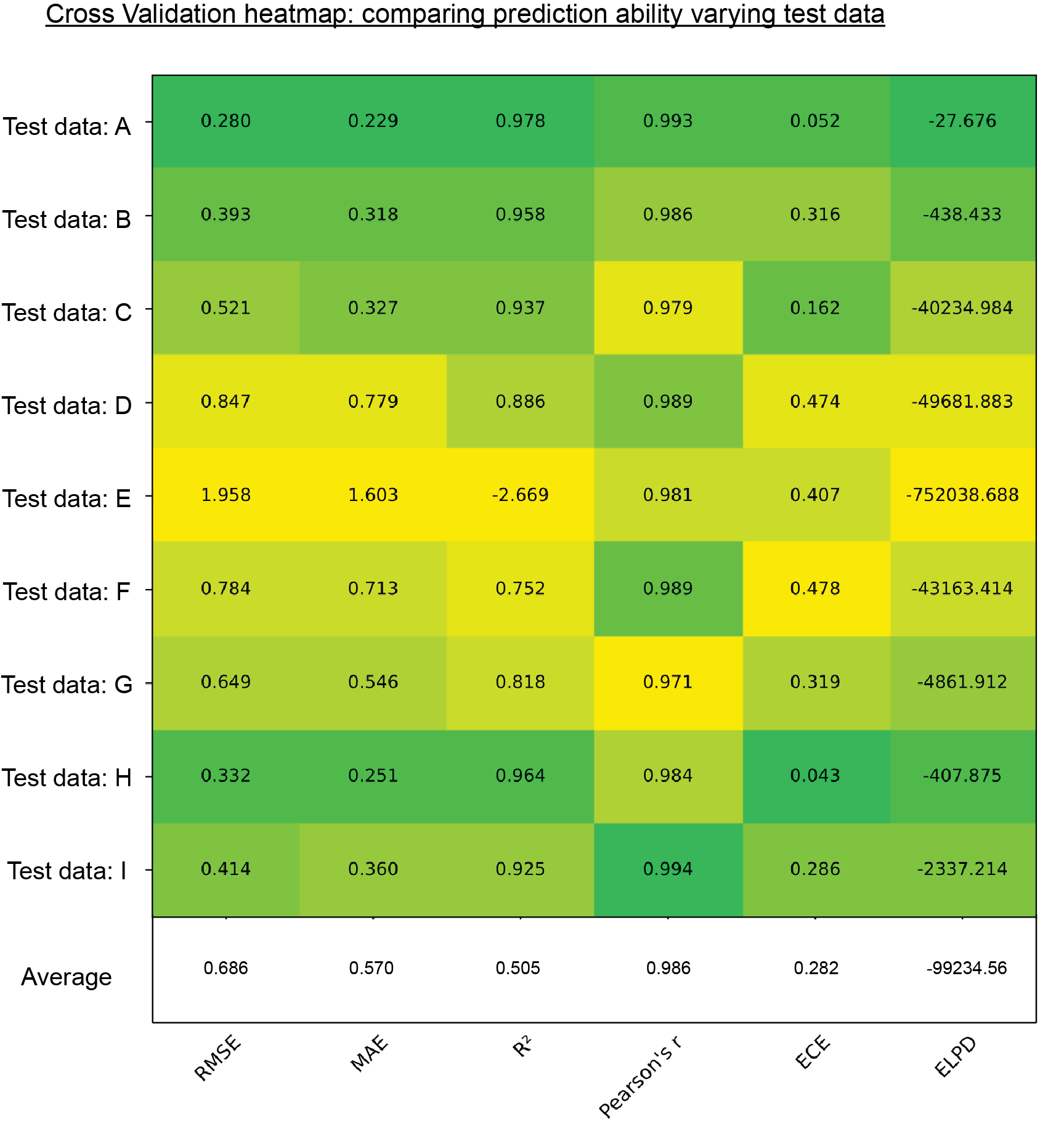}
    \caption{Leave-one-out cross validation assessment for the porous planar model. Each training run was performed while rotating out which experiment was used as the test set and therefore held out from the training. The larger error from the prediction when experiment E was the test set suggests that the data from this experiment (how the cyclic voltammogram behaved at the low total loading and 20 percent carbon support) is informative to the model and therefore difficult to predict without.}
    \label{supp-fig:cross-validation}
\end{figure}

\newpage 
\begin{figure}[htbp]
    \centering
    \includegraphics[width=0.72\textwidth]{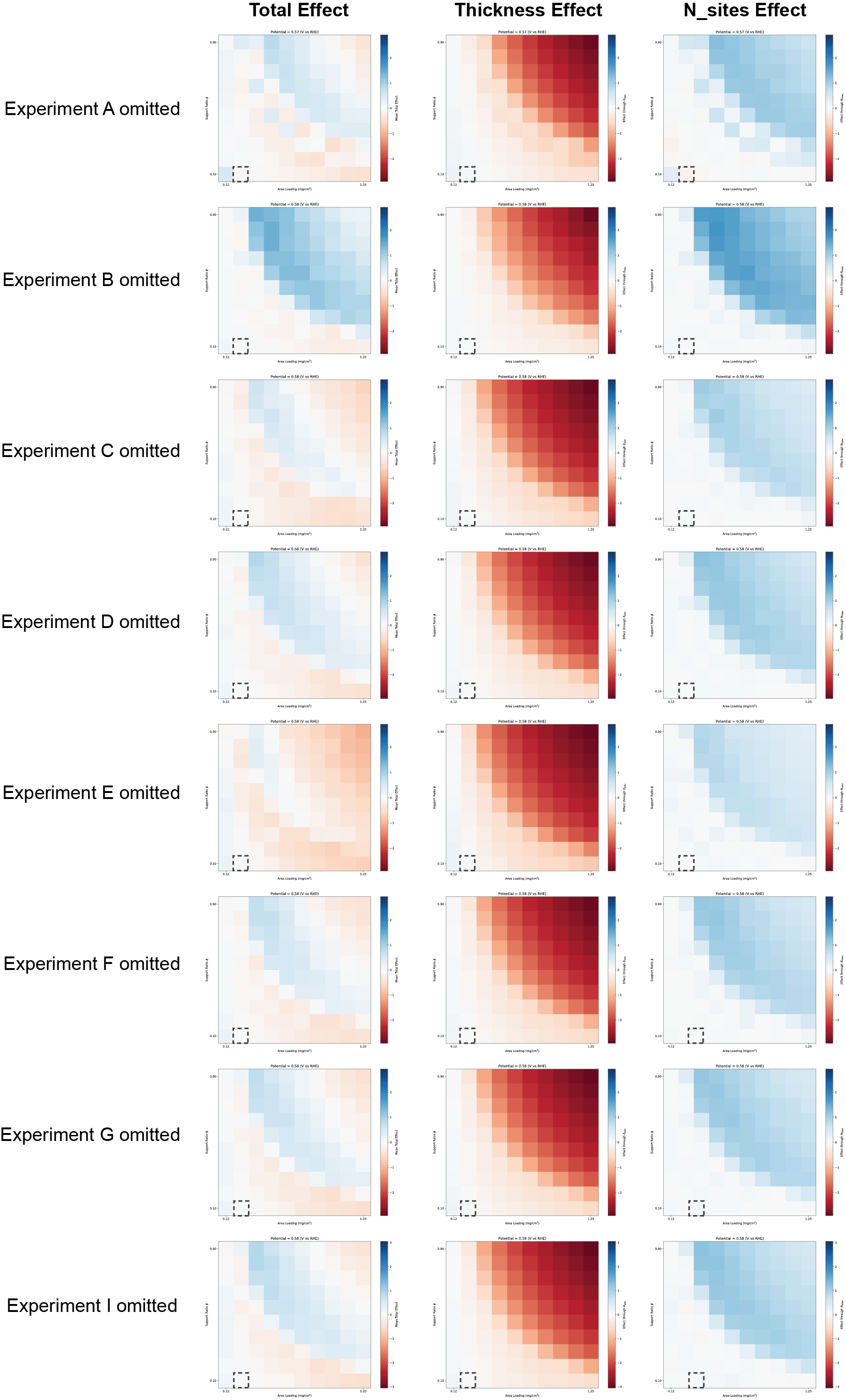}
    \caption{Mediation analysis heatmaps at 0.58V vs RHE for different training regimes where one of the experiments was fully omitted from the dataset. The trends and conclusions observed and discussed with Figures 7 and 8 are preserved even with the omission of single data points.}
    \label{supp-fig:hold-outs-mediation}
\end{figure}

\newpage
\begin{figure}[htbp]
    \centering
    \includegraphics[width=0.72\textwidth]{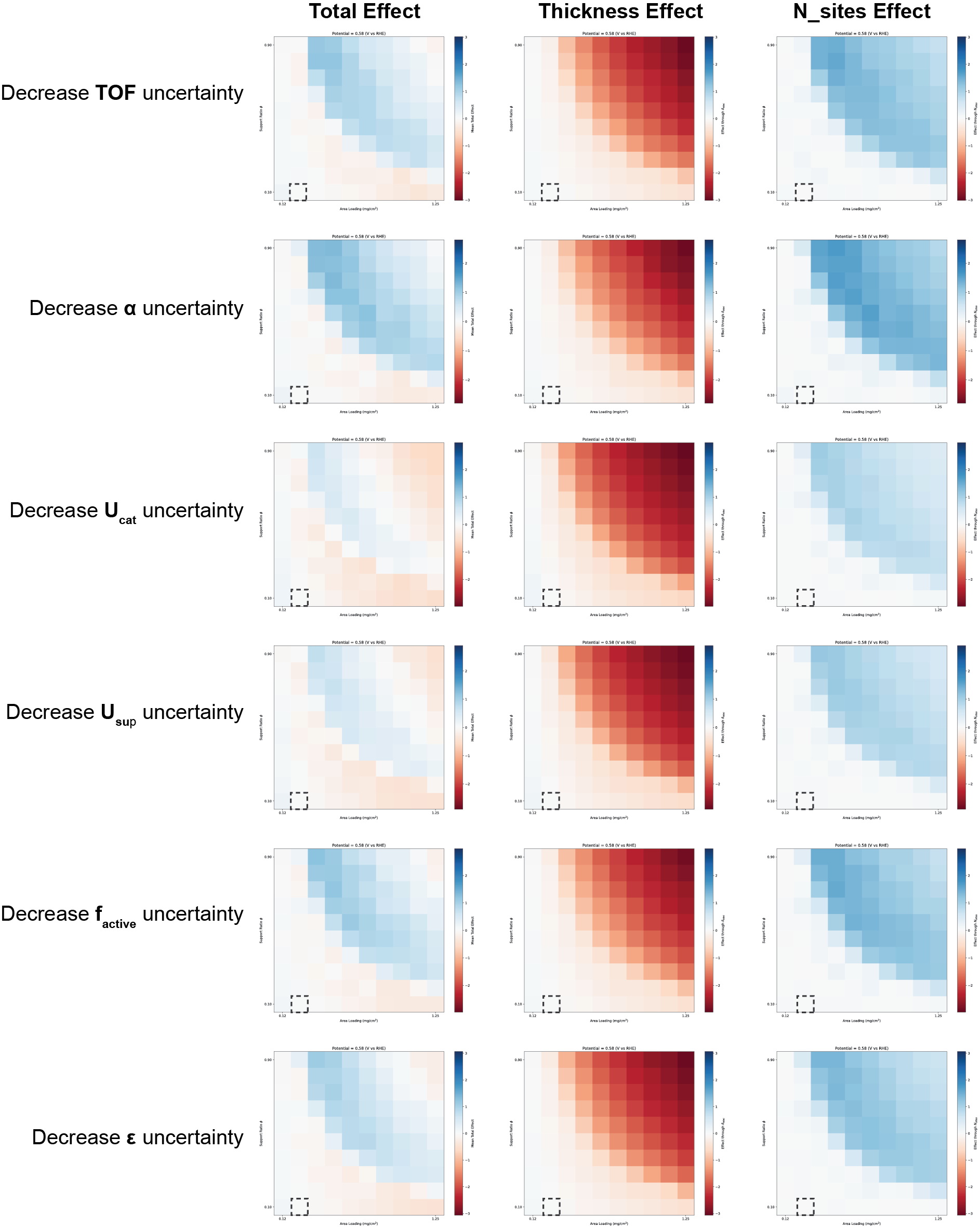}
    \caption{Mediation analysis heatmaps at 0.58V vs RHE for different training regimes where the kernel sizes are adjusted for several nodes: ToF, $\alpha$, $U_{cat}$, $U_{sup}$, $f_{act}$, $\epsilon$. Again, the trends and conclusions observed and discussed with Figures 7 and 8 are preserved even with the omission of single data points.}
    \label{supp-fig:kernels-mediation}
\end{figure}

\newpage
\begin{figure}[htbp]
    \centering
    \includegraphics[width=\textwidth]{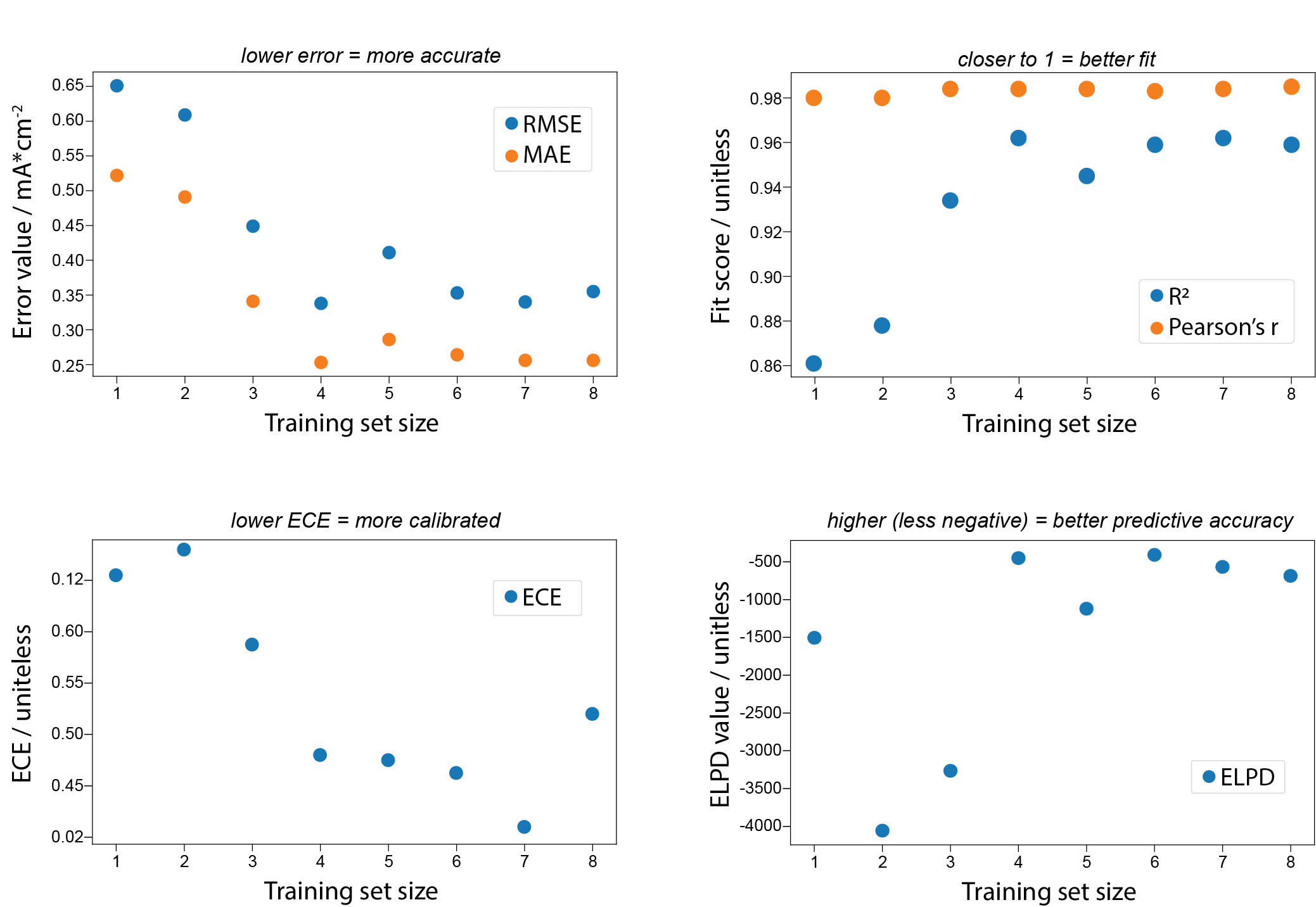}
    \caption{Learning rates are plotted for the various quantitative performance metrics used; training set size from 1-9. The sequence of experiments to add was generated randomly. Performance increases as training set size increases: decreased error, increased fit scores, and increased calibration.}
    \label{supp-fig:learning-rates}
\end{figure}

\newpage
\begin{figure}[htbp]
    \centering
    \includegraphics[width=\textwidth]{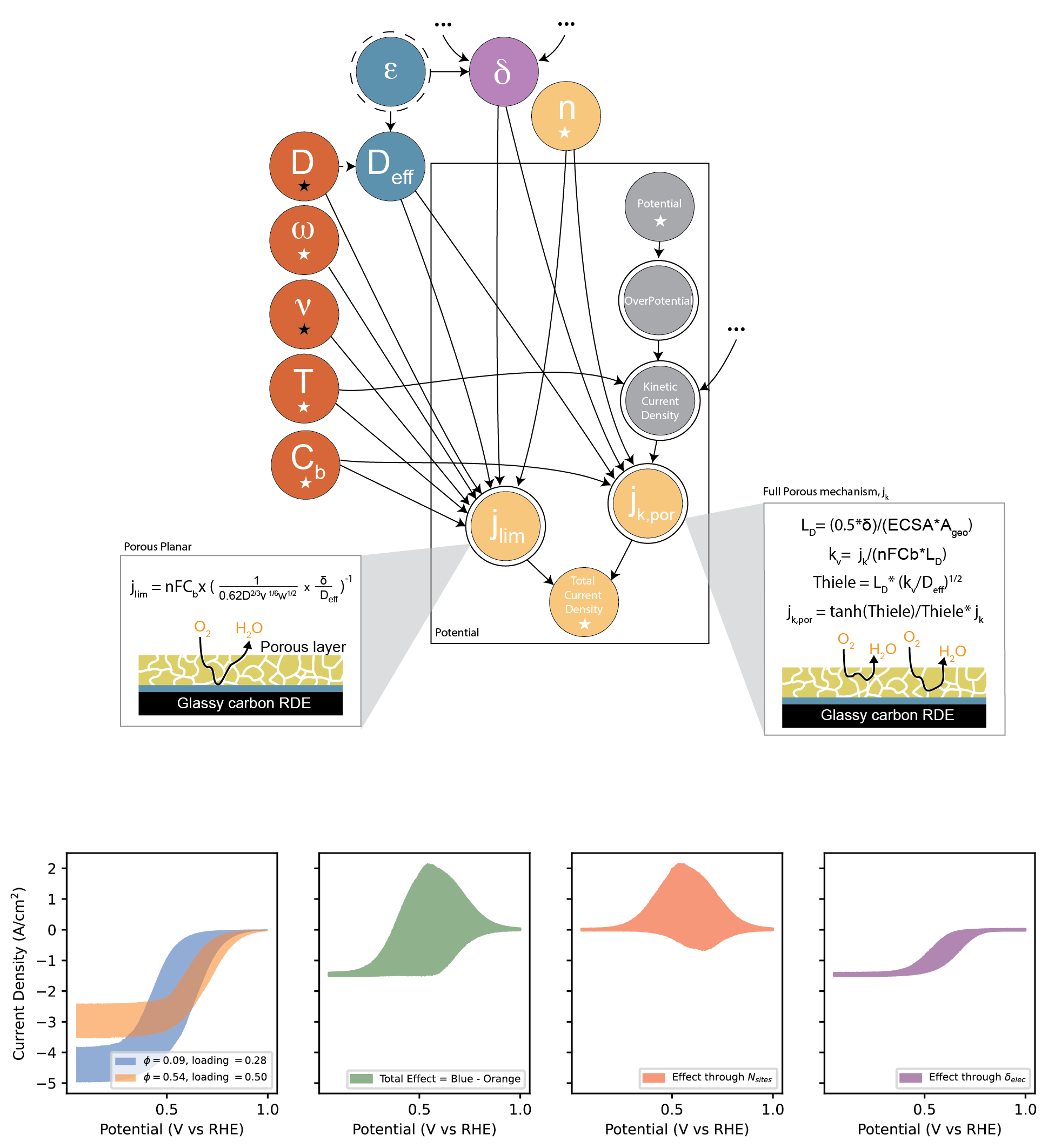}
    \caption{An updated model including a reaction-sink concentration corrected kinetic current. A portion of the SCM schematic is shown summarizing the mechanisms used to model reaction throughout the electrode thickness~\cite{dix2020critical} - the Thiele modulus is calculated to describe the relation between diffusion in the pores and reaction rate, with the characteristic diffusion length $L_D$ scaled by the electrode thickness, electrochemical surface area, and geometric surface area. The limiting current density mechanism is the same as the mechanism used in the porous planar model. A snap shot of the mediation analysis is also shown. The diffusion and reaction limited model uses the Thiele modulus to capture the relationship between reaction-rate and diffusion in a porous media. This set of equations makes the model assumptions that total reaction kinetic current is modulated by the volumetric reaction rate and characteristic diffusion length, thereby including the depletion of reactants during reaction as they diffuse through the volume of the electrode. We observe comparable mediation results to the porous planar model, however the true variation in reaction current due to concentration of species can be further improved with the inclusion of differential equations which require software updates.}
    \label{supp-fig:full-porous}
\end{figure}

\newpage
\textbf{Model improvements discussion}:  
Additional causal mechanisms connecting the support ratio to utilization factors (i.e., how support ratio directly causes the utilization factor) may exist and are absent in current models. Similarly, support ratio likely also affects the electrode porosity. To the best of our knowledge, there are no physical equations that accurately model the causes between support ratio and utilization or porosity. Modeling these causal relationships would need to leverage empirical observations from systematic studies as well as include the effects of ionomer composition/ratio in the SCM. Extensive work would need to be done validating any proposed causal relationships and applicability to the system we chose to model. For instance, future work characterizing dispersion via TEM, CO pulse adsorption, ionomer-support or ionomer-catalyst binding calculations could be included in an SCM.

Regarding selectivity, for all the models and equations used in this work, the selectivity was treated as experimental data (previously measured by Kreider et al). We propose the inclusion of the two separate reaction pathways (4e\textsuperscript{-} and 2e\textsuperscript{-}) into the SCM such that each reaction pathway has a different turnover frequency and therefore kinetic current. The kinetic currents for each pathway can therefore be used to determine the selectivity, which is then used to calculate the limiting current. In this way, the extracted selectivities measured by Kreider et al. can still be used in model training, but the two separate kinetic currents offer better fits to the sloping voltammograms in the MnSb\textsubscript{2}O\textsubscript{6} catalyst system.

Separate activity of the Vulcan carbon support would also improve the modeling efforts. This will help deconvolute the support ratio effects on the total number of active sites; mediation analyses can be performed to identify whether the increase in N sites from increased support ratio was due to increased support activity or increased catalyst utilization at some optimal ratio. 

Additionally, the porous planar SCM with the modified Levich equation was unable to represent electrochemical activity throughout a porous electrode. The modified Levich equation approximates porous electrodes as a porous layer enforcing the additional electrode thickness based mass transport limitations and electrochemical reactions occur after the reactants have diffused to the back boundary layer of the film. However, improved modeling of distributed reactions requires further SCM method development for the inclusion of the concentration profiles, Fick’s laws of diffusion, and boundary conditions. Such improvements to causally model [ordinary and partial] differential equations are a major focus of our future method development.

Furthermore, this modeling framework can be expanded to temporal evolution modeling using dynamic Bayesian networks, thus enabling causal degradation modeling in complex devices. 



\clearpage

\putbib
\end{bibunit}

\end{document}